\documentclass[%
prc,
reprint,
amsmath,amssymb,
aps,
superscriptaddress,
showpacs,
showkeys,
preprintnumbers,
nofootinbib,
]{revtex4-1}

\usepackage{graphicx}
\usepackage[utf8]{inputenc}
\usepackage[T1]{fontenc}
\usepackage{epsfig}
\usepackage{bm}
\usepackage{epstopdf}
\usepackage{dcolumn}
\usepackage{bm}
\usepackage[usenames]{color}
\usepackage{longtable} 
\usepackage{color}

\usepackage{hyperref}
\usepackage{color}
\usepackage{xfrac}
\hypersetup{pageanchor=false} 
\begin{document}

\title{Fast-timing study of \texorpdfstring{$^{\bm{81}}$}{81}Ga from the \texorpdfstring{$\bm{\beta}$}{Beta} decay of \texorpdfstring{$^{\bm{81}}$}{81}Zn}

\author{V.~Paziy}
\affiliation{Grupo de F\'{\i}sica Nuclear \& IPARCOS, Facultad de Ciencias F\'{\i}sicas, Universidad Complutense - CEI Moncloa, E-28040 Madrid, Spain}
\author{L.M.~Fraile}
\email [E-mail at: ] {lmfraile@ucm.es}
\affiliation{Grupo de F\'{\i}sica Nuclear \& IPARCOS, Facultad de Ciencias F\'{\i}sicas, Universidad Complutense - CEI Moncloa, E-28040 Madrid, Spain}
\author{H.~Mach}
\thanks{Deceased. See acknowledgments.}
\affiliation{Grupo de F\'{\i}sica Nuclear \& IPARCOS, Facultad de Ciencias F\'{\i}sicas, Universidad Complutense - CEI Moncloa, E-28040 Madrid, Spain}
\affiliation{National Centre for Nuclear Research, BP1, ul. Ho\.{z}a 69, 00-681, Warsaw, Poland}
\author{B.~Olaizola}
\thanks{Present address: TRIUMF, 4004 Wesbrook Mall, Vancouver, British Columbia V6T 2A3, Canada}
\affiliation{Grupo de F\'{\i}sica Nuclear \& IPARCOS, Facultad de Ciencias F\'{\i}sicas, Universidad Complutense - CEI Moncloa, E-28040 Madrid, Spain}
\author{G.S.~Simpson}
\affiliation{LPSC, Universit\'{e} Joseph Fourier Grenoble 1, CNRS/IN2P3, Institut National Polytechnique de Grenoble, F-38026 Grenoble Cedex, France}
\author{A.~Aprahamian}
\affiliation{Department of Physics, University of Notre Dame, Notre Dame, Indiana 46556, USA}
\author{C.~Bernards}
\affiliation{Institut f\"{u}r Kernphysik, Universit\"{a}t zu K\"{o}ln, Z\"{u}lpicher Strasse 77, D-50937 K\"{o}ln, Germany}
\affiliation{Wright Nuclear Structure Laboratory, Yale University, New Haven, Connecticut 06520, USA}
\author{J.A.~Briz}
\affiliation{Instituto de Estructura de la Materia, CSIC, 28006 Madrid, Spain}
\author{B.~Bucher}
\affiliation{Idaho National Laboratory, Idaho Falls, Idaho 83415, USA}
\author{C.~J.~Chiara}
\thanks{Present address: U.S. Army Research Laboratory, Adelphi, Maryland 20783, USA}
\affiliation{Department of Chemistry and Biochemistry, University of Maryland, College Park, Maryland 20742, USA}
\affiliation{Physics Division, Argonne National Laboratory, Argonne, Illinois 60439, USA}
\author{Z.~Dlouh\'y}
\thanks{Deceased.}
\affiliation{Nuclear Physics Institute of the AS CR, Z-25068, \v{R}e\v{z}, Czech Republic}
\author{I.~Gheorghe}
\affiliation{\textquotedbl{Horia Hulubei}\textquotedbl National Institute for Physics and Nuclear Engineering, R-77125 Bucharest-Magurele, Romania}
\author{D.~Ghi\c{t}\v{a}}
\affiliation{\textquotedbl{Horia Hulubei}\textquotedbl National Institute for Physics and Nuclear Engineering, R-77125 Bucharest-Magurele, Romania}
\author{P.~Hoff}
\affiliation{Department of Chemistry, University of Oslo, P.O. Box 1033 Blindern, N-0315 Oslo, Norway}
\author{J.~Jolie}
\affiliation{Institut f\"{u}r Kernphysik, Universit\"{a}t zu K\"{o}ln, Z\"{u}lpicher Strasse 77, D-50937 K\"{o}ln, Germany}
\author{U.~K\"oster}
\affiliation{Institut Laue-Langevin, 71 avenue des Martyrs, 38042 Grenoble Cedex 9, France}
\author{W.~Kurcewicz}
\affiliation{Faculty of Physics, University of Warsaw, PL 02-093 Warsaw, Poland}
\author{R.~Lic\v{a}}
\affiliation{\textquotedbl{Horia Hulubei}\textquotedbl National Institute for Physics and Nuclear Engineering, R-77125 Bucharest-Magurele, Romania}
\author{N.~M\v{a}rginean}
\affiliation{\textquotedbl{Horia Hulubei}\textquotedbl National Institute for Physics and Nuclear Engineering, R-77125 Bucharest-Magurele, Romania}
\author{R.~M\v{a}rginean}
\affiliation{\textquotedbl{Horia Hulubei}\textquotedbl National Institute for Physics and Nuclear Engineering, R-77125 Bucharest-Magurele, Romania}
\author{J.-M.~R\'egis}
\affiliation{Institut f\"{u}r Kernphysik, Universit\"{a}t zu K\"{o}ln, Z\"{u}lpicher Strasse 77, D-50937 K\"{o}ln, Germany}
\author{M.~Rudigier}
\thanks{Present address: Institut f\"ur Kernphysik, Technische Universit\"at Darmstadt, D-64289, Darmstadt, Germany}
\affiliation{Institut f\"{u}r Kernphysik, Universit\"{a}t zu K\"{o}ln, Z\"{u}lpicher Strasse 77, D-50937 K\"{o}ln, Germany}
\author{T.~Sava}
\affiliation{\textquotedbl{Horia Hulubei}\textquotedbl National Institute for Physics and Nuclear Engineering, R-77125 Bucharest-Magurele, Romania}
\author{M.~St\v{a}noiu}
\affiliation{\textquotedbl{Horia Hulubei}\textquotedbl National Institute for Physics and Nuclear Engineering, R-77125 Bucharest-Magurele, Romania}
\author{L.~Stroe}
\affiliation{\textquotedbl{Horia Hulubei}\textquotedbl National Institute for Physics and Nuclear Engineering, R-77125 Bucharest-Magurele, Romania}
\author{W.B.~Walters}
\affiliation{Department of Chemistry and Biochemistry, University of Maryland, College Park, Maryland 20742, USA}

\begin{abstract}

The $\beta^{-}$ decay of $^{81}$Zn to the neutron magic $N=50$ nucleus $^{81}$Ga, with only three valence protons with respect to $^{78}$Ni, was investigated. 
The study was performed at the ISOLDE facility at CERN by means of $\gamma$ spectroscopy. 
The $^{81}$Zn half-life was determined to be $T_{1/2}=290(4)$ ms while the $\beta$-delayed neutron emission probability was measured as $P_n=23(4)\%$. The analysis of the $\beta$-gated $\gamma$-ray singles and $\gamma$-$\gamma$ coincidences from the decay of $^{81}$Zn provides 47 new levels and 70 new transitions in $^{81}$Ga. The $\beta^-$$n$ decay of $^{81}$Zn was observed and a new decay scheme into the odd-odd $^{80}$Ga nucleus was established. The half-lives of the first and second excited states of $^{81}$Ga were measured via the fast-timing method using LaBr$_3$(Ce) detectors. The level scheme and transition rates are compared to large-scale shell-model calculations. The low-lying structure of $^{81}$Ga is interpreted in terms of the coupling of the three valence protons outside the doubly-magic $^{78}$Ni core.

\end{abstract}


\maketitle

\section{Introduction}

Modifications to the standard ordering of the single-particle energies have been observed in exotic nuclei with a large disparity in proton and neutron numbers. They give rise to the disappearance of the conventional magic numbers and the appearance of new shell gaps. The understanding of the underlying physics driving such modifications is one of the main subjects of modern nuclear-structure studies. 
It is recognized that monopole shifts are responsible for the evolution of shell structure far off stability, but the effect of the different components of 
the monopole interaction is still the subject of investigation \cite{otsuka10}, since it is not simple to disentangle them from the experimental information. 
This is mainly due to the fact that effective single-particle energies (ESPEs) cannot be directly measured, and that single-particle and collective
effects arising from residual interactions are intertwined. The central term of the monopole interaction seems to be responsible for the evolution of
ESPEs, while the tensor term plays a leading role in the splitting of spin-orbit partners. 

The regions in the immediate vicinity of exotic doubly-magic nuclei are key for mapping the single-particle degrees of freedom around closed cores. The evolution of the proton-neutron interaction arising from the tensor force and the role of neutron excitations across neutron shell gaps can be studied in these nuclei. Relevant ingredients to theoretical models can also be obtained. Two unexplored areas in the table of nuclides still remain: around the doubly-magic $^{78}$Ni and in the vicinity of $^{100}$Sn.

Although $^{78}$Ni, with 28 protons and 50 neutrons ($Z=28$, $N=50$), is located 14 neutrons off the stability line, it is expected to be a
doubly-magic nucleus due to the robust shell gaps arising from the spin-orbit splitting both for protons ($\pi f_{7/2}-\pi f_{5/2}$) and
neutrons ($\nu g_{9/2}-\nu g_{7/2}$). Evidence for strong $Z=28$ and $N=50$ shell closures in $^{78}$Ni has been recently obtained 
\cite{taniuchi19}. In this work the role of collective effects in such an exotic nucleus, which were subject of debate, has also been highlighted.
It is therefore of the greatest interest to understand its structure and that of nuclei around the $Z=28$, $N=50$ double shell closure. 

The first evidence for the existence of $^{78}$Ni came from \cite{eng95}.
Afterwards its half-life was reported \cite{hosmer10} to be $T_{1/2}=100^{+100}_{-60}$ ms and more recently $T_{1/2}=122\pm5$ ms \cite{xu14}. The latter value does point towards the magic character of $^{78}$Ni. 
Theoretical calculations predicted the first excited state energy above 2 MeV \cite{nowacki16,hagen16}, which would also be consistent with a doubly-magic character. 
Only recently in-beam $\gamma$-ray spectroscopy of the elusive $^{78}$Ni was performed \cite{taniuchi19}. The experimental results together with theoretical calculations \cite{taniuchi19} confirm the magic nature of $^{78}$Ni, but suggest competing spherical and deformed configurations in the region, and predict the breakdown of the $Z=28$ shell closure towards heavier nickel isotopes. In this context, mapping the $Z=28$ isotopes and the $N=50$ isotones is of great interest. 
Monopole drifts have been observed in neighboring $Z=29$ Cu isotopes leading to the modification of ground-state configurations \cite{flanagan09,ichikawa19}, which may also point to a weakening of the $Z=28$ gap. 

The strength of the $N=50$ neutron shell gap and the proton structure close to $^{78}$Ni can be obtained from the $N=50$ isotones, and
in particular from the odd-proton neighbors $^{79}$Cu and $^{81}$Ga. The nucleus $^{79}$Cu was not reachable until very recently, when the
first spectroscopic study was reported \cite{olivier17} and its mass was precisely measured \cite{welker17}. From these studies the magicity 
of $^{78}$Ni and the persistence of the $Z=28$ gap is confirmed. In this way $^{79}$Cu can be described as a valence proton coupled to the 
$^{78}$Ni core. A spin-parity of 5/2$^-$ is suggested for its ground state (g.s), while the 3/2$^-$ first-excited state is proposed at a high energy
of 656 keV \cite{olivier17}. The lowering of the 5/2$^-$ state and eventual inversion with the 3/2$^-$ is shown for the Cu isotopic chain by
recent Monte Carlo shell-model (MCSM) calculations \cite{ichikawa19}. The $^{79}$Cu results are consistent with the description of $^{80}$Zn, two 
protons above $^{78}$Ni, in terms of two-proton configurations on top of the $^{78}$Ni core \cite{shiga16}, which also confirm the persistence 
of the $N=50$ shell closure. 

The next odd $N=50$ isotope, $^{81}$Ga, is the subject of this paper. With three protons outside $^{78}$Ni, it provides important information about
proton single-particle configurations and on the strength of the $N=50$ shell closure when the number of protons increase. In our study we have produced $^{81}$Zn isotopes at ISOLDE, CERN to populate $^{81}$Ga in $\beta^-$ decay. We have used $\gamma$-ray spectroscopy to greatly extend the known level scheme, and the Advanced Time Delayed (\textquotedbl{fast-timing}\textquotedbl) $\beta\gamma\gamma$(t) method \cite{mach89,mosz89} to measure excited level lifetimes, and deduce transition probabilities, which provide more stringent tests of the theoretical models and will help interpret the structure of $^{81}$Ga. 



\section{Previous knowledge of \texorpdfstring{$^{\bm{81}}$}{81}G\lowercase{a}}

The first studies of the decay of $^{81}$Zn were performed in 1991 at the ISOLDE, CERN facility by Kratz \textit{et al.} \cite{kra91}. The half-life and the $\beta$-delayed neutron emission probability were investigated, the reported values being $T_{1/2}=290(50)$ ms and $P_n=7.5(30)\%$.
Later, two $\gamma$ transitions of 351 and 452 keV were identified as belonging to the decay of $^{81}$Zn to $^{81}$Ga by Verney \textit{et al.} \cite{ver04} at PARRNe, and by K\"{o}ster \textit{et al.} \cite{koester05} at ISOLDE. In the latter measurement, due to the notable $^{80}$Ga activity present in the decay of $^{81}$Zn, a lower limit of 10\% for the $P_n$ value was suggested. Theoretical calculations \cite{borzov05} predicted $P_n=13\%$. Measurements performed at the NSCL and published in 2010 by Hosmer \textit{et al.} \cite{hosmer10} proposed a considerably longer half-life of $474^{+93}_{-83}$ ms and a higher $P_n=30(13)\%$ value for $^{81}$Zn.

The $\beta^-$ decay of $^{81}$Zn was studied again at the PARRNe mass separator in \cite{verney07}. The statistics obtained in this experiment allowed for the 351.1-keV transition to be attributed to $^{81}$Ga due to the new $^{81}$Zn half-life value of $391(65)$ ms. The existence of the second excited state at 802.8 keV was confirmed by the observation of a 451.7-keV $\gamma$ ray in coincidence with the 351.1-keV line. The first excited state was defined by the 351.1-keV transition based on $\gamma$-intensity considerations. A third, weak transition was detected at 1621.6 keV and a tentative state of the same energy was added to the level scheme. Spin assignments for the ground, first-excited and second-excited states of $^{81}$Ga were tentatively proposed to be $(5/2^{-})$, $(3/2^{-})$, and $(3/2^{-})$, respectively, based on shell-model calculations and proton single-particle states. The authors suggested $(1/2^{+})$ spin-parity for the $^{81}$Zn ground state. Magnetic-moment measurements performed at ISOLDE \cite{cheal10} yielded a $^{81}$Ga ground-state spin-parity value of 5/2$^-$, confirming the earlier tentative assignment. 

A study of the $^{81}$Ga structure was performed at LNL via heavy-ion multi-nucleon transfer \cite{sahin12}. Several $\gamma$ rays were attributed to $^{81}$Ga, and specifically a 1236-keV transition connecting a state of the same energy to the g.s., which were assigned $(9/2^{-})$ and $(5/2^{-})$ spin-parities, respectively. A new measurement of the $yrast$ states of $^{81}$Ga populated in fission \cite{dudouet19} contradicts this assignment, since none of the $\gamma$ rays reported in \cite{sahin12} could be confirmed. Instead the $(9/2^{-})$ is observed at 1340.7 keV. The 1398.5-keV and 1952.2-keV levels are assigned $(7/2^{-})$ and $(11/2^{-})$ spin-parity, respectively \cite{dudouet19}. The two later states are also observed by in-beam spectroscopy in knockout reactions at RIBF \cite{olivier17PhD}. In spite of the large spin difference, the indirect population of these higher spin states in $^{81}$Ga from the $\beta$ decay of ($1/2^{+}$) $^{81}$Zn ground state should be possible. 

The most recent data of the $\beta$ decay of $^{81}$Zn comes from HRIBF at ORNL. The results were published in 2010 by Padgett \textit{et al.} \cite{pad10}. The decay scheme showed six new energy levels and nine new $\gamma$ transitions in addition to the previously available ones. Transitions of 451, 916, 1107, 1585 and 2358 keV were observed in coincidence with the strongest 351-keV peak. Four other $\gamma$ rays, only observed in the singles $\gamma$ spectrum, with energies of 1458, 1936, 4294 and 4880 keV were placed directly feeding the ground state. A new value of $304(13)$ ms for the $^{81}$Zn half-life was established and a $\beta$-delayed neutron branch of 12(4)\% was determined using the 1083-keV transition in $^{80}$Ge. In this work, a spin-parity assignment of $(5/2^{+})$ for the ground state of $^{81}$Zn, different from the earlier value, was proposed, based on the $J^\pi=5/2^-$ $^{81}$Ga g.s. \cite{cheal10} and the $\beta$-feeding pattern. The first and the second excited states both received a tentative $J^\pi=(3/2^-)$ assignment. 

From the existing works, the structure of $^{81}$Ga is interpreted as arising from the coupling of valence protons in the $fp$ shell, leading to negative-parity states at low excitation energy. Positive-parity states, requiring the excitation of a proton to the $g_{9/2}$ orbit across $Z=40$, or particle-hole excitations of the $^{78}$Ni core, appear at energies above 4 MeV. The high excitation energy points towards a robust $N=50$ neutron shell closure, in agreement with the recently observed $^{79}$Cu structure \cite{olivier17}.

\section{Experimental setup and data analysis}
\label{sec:experiment}

The present experiment was performed at the ISOLDE, CERN facility in the framework of a systematic fast-timing investigation of neutron-rich nuclei populated following the decay of Zn isotopes {\cite{RAZ14,paziy16PhD,vedia2017b}}. The selectivity and efficiency for the production of Zn ion beams had been previously optimized \cite{koester08} in order to enhance the beam purity for $^{77-82}$Zn ions. Proton pulses with an average charge of 5~$\mu$C and 1.4-GeV energy, coming from the PS-Booster in intervals of multiples of 1.2~s, were converted into fast neutrons {\cite{koester02}} that impinged onto a hot $\sim$2000 $^o$C UC$_2$/graphite target, inducing fission reactions. The thermally extracted products were guided through a temperature-controlled quartz glass transfer line 
{\cite{bouquerel07}} into a W ionizer where selective ionization was performed by the ISOLDE Resonance Ionization Laser Ion Source (RILIS) \cite{RILIS}. The single-charged $A=81$ ions were mass separated by the magnetic high resolution mass separator (HRS), accelerated to 60 keV, and directed to the experimental setup.

\begin{figure}
	\centerline{\includegraphics[scale=0.25]{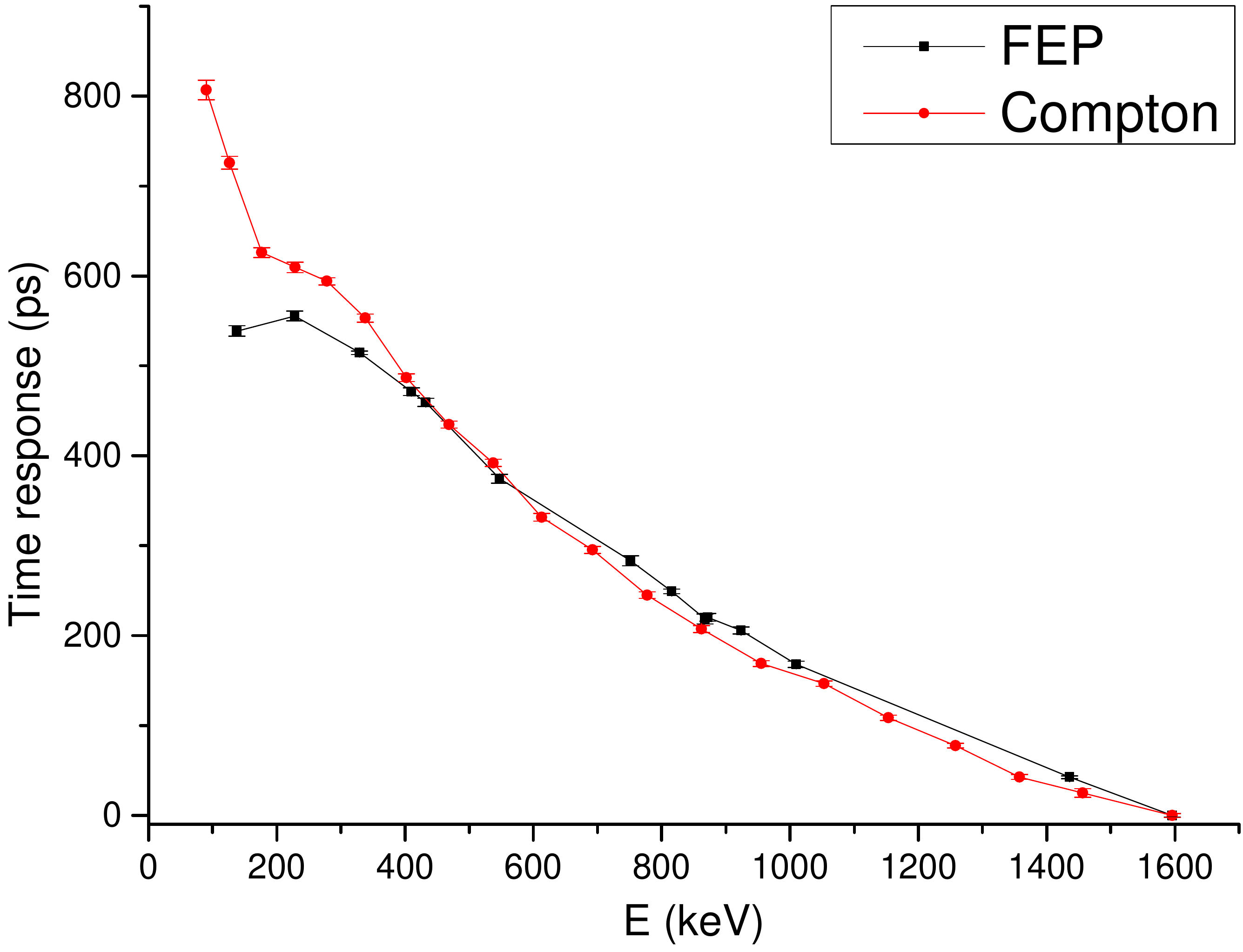}}
	\caption{\label{Compton-Prompt}Relative Compton response (red) and full-energy peak (FEP) prompt curve (black) of one of the LaBr$_{3}$(Ce) crystals used in our experimental setup. The calibration is obtained with an $A=140$ source.}
\end{figure}

The mass-separated $^{81}$Zn nuclei were continuously collected on an aluminum stopper foil, creating a saturated source. The estimated yield of $^{81}$Zn was 600 ions/$\mu$C. 
Since Rb atoms partially survived the quartz transfer line selection and were surface ionized on the walls of the ionizer, the long-lived $T_{1/2}=4.57$ h contaminant $^{81}$Rb was present in the beam, with about five times higher production than $^{81}$Zn, but much lower activity during the data taking. An electrostatic deflector (beam gate) blocking the delivery of ions to the experimental station, was used to avoid the accumulation of long-lived $^{81}$Rb activity coming from the target long after most of the $^{81}$Zn had been released. For the mass 81 experiment, the beam gate was closed 600~ms after the beam impinged on the target, and the collected species were allowed to decay out. 

\begin{figure*}[ht!]
	\includegraphics[scale=0.35]{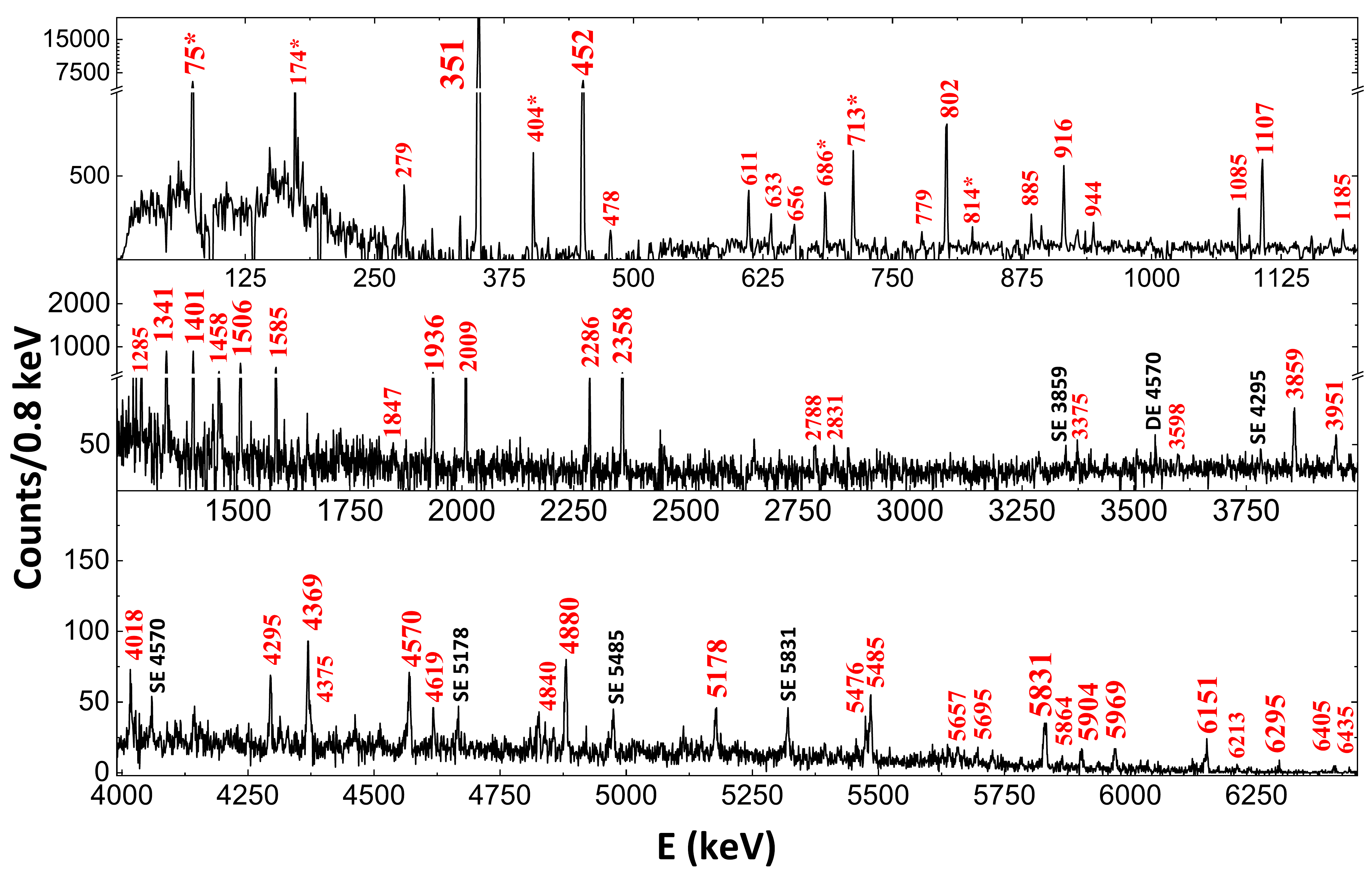}
	\caption{\label{beta-gamma-subtracted} The $\beta$-gated $\gamma$-ray singles spectrum obtained following the decay of $^{81}$Zn, after subtraction of the long-lived activity. The transitions in $^{81}$Ga are labeled by their energies. Some transitions from the $\beta^{-}$n decay of $^{81}$Zn to $^{80}$Ga are marked with asterisks.}
\end{figure*}

The experimental setup included two HPGe detectors, two LaBr$_{3}$(Ce) detectors, and an NE111A plastic scintillator for $\beta$-particle detection, very close to the beam deposition point. In particular, the 3-mm-thick plastic scintillator was located less than 1 mm away from the stopper foil in order to maximize the detection efficiency. This thin detector assures ultra-fast and uniform time response independent of the incident $\beta$ energy. 
The germanium detectors were used for the detection of $\gamma$ radiation in the range of 30 to 7000 keV; their energy resolution was 2.0 keV at $^{60}$Co energies. Coincidences with the $\beta$ detector were used for $\gamma$-ray background suppression, and $\gamma$-$\gamma$ coincidences between the HPGe detectors to determine the decay scheme. For the lifetime measurements of the excited states in the tens of picoseconds to nanosecond range, fast-response inorganic LaBr$_{3}$(Ce) crystals with the shape of truncated cones \cite{vedia2017} were mounted almost perpendicularly to the germanium detectors. These scintillator crystals have a fast-decay component that makes it possible to achieve very good time resolution while maintaining acceptable energy resolution \cite{vedia2015,vedia2017}. Each crystal was mounted onto a Photonis XP20D0 fast-response 2-inch photomultiplier tube (PMT), optimized to give fast time response at the cost of lower gain.

The signals from all the detectors were processed by a digital data acquisition (DAQ) system composed of four Pixie-4 Digital Gamma Finder cards, specially designed for $\gamma$-ray spectroscopy \cite{pixie4}. For the energy analysis, the HPGe signals from the preamplifier were fed into the DAQ, while the much faster scintillator signals taken from the last PMT dynodes were shaped before they were sent to the digital system. The PMT anode signals from the scintillator detectors were used for fast timing. The signals were processed by analog Constant Fraction Discriminators (CFD), and then sent to Time to Amplitude Converter (TAC) modules to measure the time difference between the $\beta$ start detector and the two $\gamma$ scintillators, which acted as stop detectors. Additionally, two more TACs were included to record time differences between the fast $\beta$ and the slower HPGe detectors.
Logic signals related to the beam parameters were also recorded including the time of proton impact on target which triggers the production and release of Zn ions out of the target. These triggered beam pulses define the starting time for the $^{81}$Zn accumulation and were used to rule out the long half-life contaminants by setting time gates with this signal as a reference. The Pixie-4 system is configured to write data in a triggerless mode. Coincident events were constructed off-line in order to correlate the time differences, the detector energies, and the other relevant running parameters.

For the data analysis, a time gate starting 50 ms after the proton impact and ending 1200 ms after it was adopted, which minimizes the presence of long-lived daughter activity in the $A=81$ data. Coincidence with $\beta$ particles was imposed to suppress the background contributions. 
The energy spectra contain $\gamma$ lines from the $^{81}$Zn decay chain and also a negligible fraction of contaminant lines from the $\beta^+$ decay of $^{81}$Rb to $^{81}$Kr. The strongest line of this decay (446 keV) was around 4\% as intense as the 351-keV transition of $^{81}$Ga. In addition, the subtraction of the long-lived activity (using a delayed time window after proton impact) provides a clean energy spectrum containing $\gamma$ rays from the $\beta$ decay of $^{81}$Zn, including the $\beta$-delayed neutron emission branch. 
The $\gamma$ rays from the decay can be assigned to de-excite energy levels in the $^{81}$Ga and $^{80}$Ga nuclei and their daughters.

In the first $\sim$50 ms after proton impact on the target, neutron-capture $\gamma$ rays are observed in the HPGe spectra. This is due to neutrons that escape the converter in the target area, thermalize and reach the measurement station. These capture lines were used for high energy calibration of the HPGe detectors up to 7 MeV, together with sources of $^{133}$Ba, $^{138}$Cs, $^{140}$Ba and $^{152}$Eu for the energy and efficiency calibrations. 

Excited-state lifetimes have been measured using the Advanced Time-Delayed $\beta\gamma\gamma$(t) fast-timing method \cite{mach89,mosz89,mach91}. Coincidences between the fast-response plastic scintillator and the LaBr$_{3}$(Ce) crystals were used. The method consists of the use of triple $\beta\gamma\gamma$ coincident events. The $\beta$-HPGe-HPGe coincidences allow the decay branches to be identified whereas the $\beta$-HPGe-LaBr$_{3}$(Ce) events make it possible to measure the lifetimes down to the tens of picoseconds range. The decay path is selected with a gate on the HPGe detector, whereas the lifetime is obtained from the time difference between the $\beta$ plastic scintillator and the LaBr$_3$(Ce) $\gamma$ signal, which start and stop a time-to-amplitude converter (TAC), respectively. 
With a FWHM time resolution of the LaBr$_{3}$(Ce) detectors of 110 ps for the $^{60}$Co full energy peaks \cite{vedia2017} and the very fast time response of the $\beta$ plastic scintillator below 50 ps, the $\beta$-LaBr$_{3}$(Ce) time distribution for prompt transitions, typically quasi-Gaussian, has a FWHM of 
120~ps. Half-lives longer than about 60 ps will appear as a slope on the delayed part of the time spectrum. The lifetime can be extracted by the de-convolution of the slope of the time spectrum from the prompt time distribution. Shorter half-lives, down to tens of ps, are obtained by the centroid shift of the time distribution with respect to the time distribution of a prompt transition of the same energy \cite{mach89}.

The application of the centroid shift method requires the use of calibration curves for the time response as a function of energy, both for the full-energy peaks (FEP) and Compton events. For the FEP prompt response curve we have used peaks from a $^{140}$Ba/$^{140}$La calibration source, primarily from excited states of $^{140}$Ce with known half-lives {\cite{mach95}}, including both the correction by the Compton curve and the level lifetime. Both curves are plotted in Fig.~\ref{Compton-Prompt}. The Compton response curve has been constructed with the time response of Compton events arising from the 1596-keV $\gamma$ transition from $^{140}$Ce. The time response curves have a smooth behavior versus energy, and they are very similar for $\gamma$-ray energies above 400 keV. At lower energies the curves differ due to the physics of the interaction \cite{mach91}, especially in the region of backscatter and X-ray events. 

Peak and background centroid corrections are made separately following their respective walk curves. Normally the FEPs sit on background arising mainly from Compton events coming from transitions with higher energies. The time delay originating from the background component is corrected according to the peak-to-background ratio with the help of the Compton correction curve \cite{mach91}. The resulting centroid of the FEP time distribution is then compared with the baseline given by the FEP correction curve. Any delay relative to the curve is then due to the lifetime carried by the transition giving rise to the FEP and can be related to levels in the nuclide of interest. In addition, the timing analysis includes standard corrections for the very small dependence of the $\beta$ time response with energy, and, if needed, for small electronics drifts during the measurement. 

\section{Results}

The significantly higher statistics obtained in our experiment compared to previous works is illustrated by the spectrum shown in Fig.~\ref{beta-gamma-subtracted}. Transitions up to 6.5 MeV in energy are observed, along with the strongest transitions at 351.1 and 451.6 keV. More than 20 $\gamma$ decays with sizeable intensity are detected beyond 4 MeV. 

\begin{figure}
	\centerline{\includegraphics[width=\columnwidth]{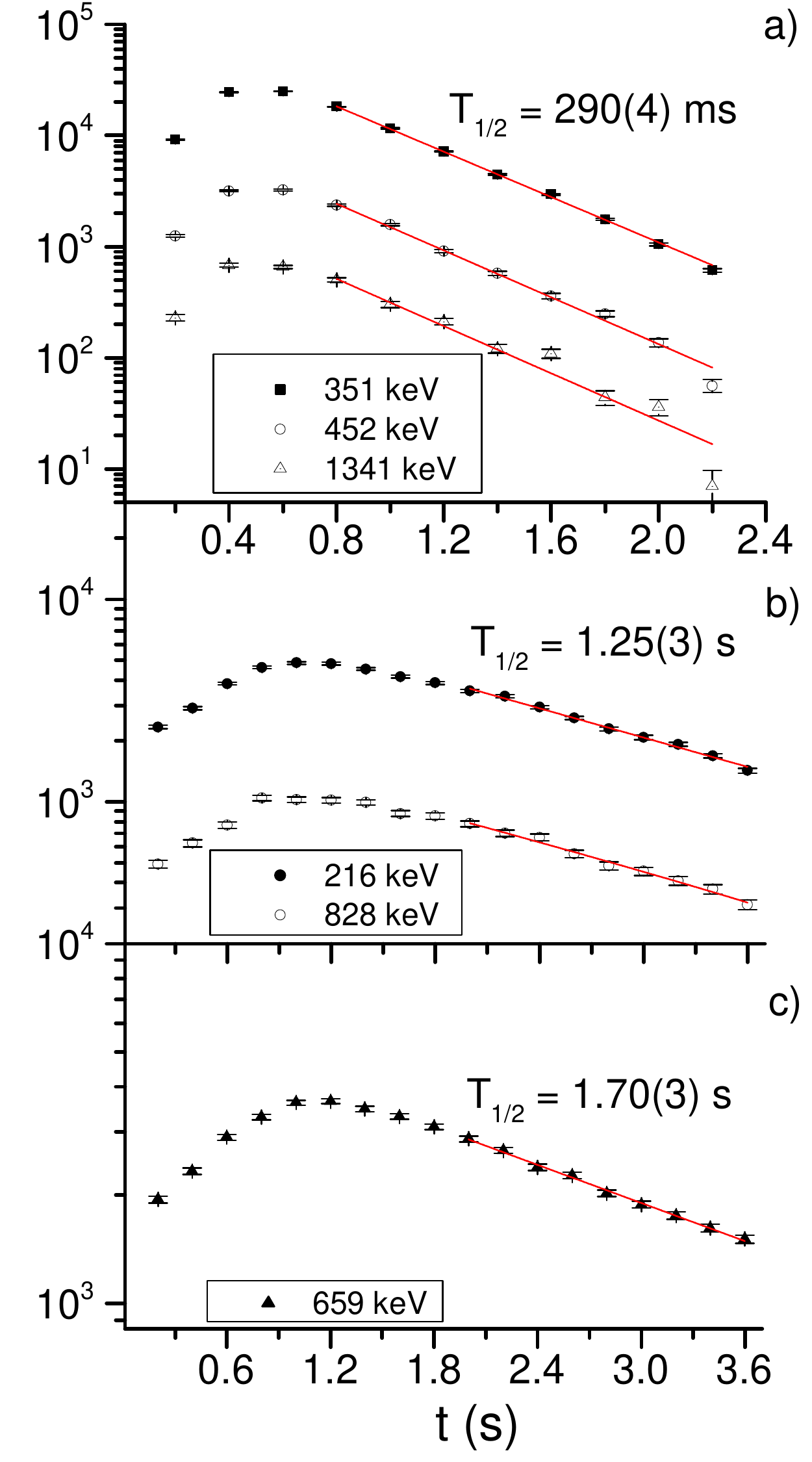}}
	\caption{\label{half_lives} Ground-state half-lives measured in this work. \\
(a) $^{81}$Zn half-life obtained from three of the strongest $^{81}$Ga transitions at 351, 452 and 1341 keV.
(b) Measurement of the $^{81}$Ga half-life by gating on the 216- and 828-keV transitions in $^{81}$Ge.
(c) Apparent $^{80}$Ga half-life (combined ground state and 22-keV isomer).
}
\end{figure}

To obtain the $^{81}$Ga half-life, the 216- and 828-keV transitions in $^{81}$Ge \cite{hoff81} were used. An exponential fit was employed by limiting the lower time boundary to 2000 ms after proton impact, which corresponds to 6.9 half-lives of $^{81}$Zn, when less than 1\% remains. The fitted slope leads to $T_{1/2}=1.25(3)$ s for $^{81}$Ga as depicted in Fig.~\ref{half_lives}, consistent with the literature value of 1.217(5)~s \cite{NDS81}.

Finally, gating on the 659-keV transition which de-excites the $2^{+}$ 659-keV level in $^{80}$Ge, we get the apparent $^{80}$Ga half-life, where a 22-keV $3^{-}$ isomer has been identified above the $6^{-}$ ground state \cite{RAZ14}. The half-lives of these states were previously measured as 1.3(2) and 1.9(1)~s, respectively, in the $\beta$-decay experiment described in \cite{ver13}. According to the level scheme from Fig.~5 of \cite{ver13}, the 659-keV state is $\beta$-fed directly from the low-spin isomer while the high-spin isomer populates it via the 1083-keV $\gamma$ ray that de-excites the 1743-keV level. Therefore, the time since proton impact spectrum gated by the 659-keV will contain the contribution of half-lives from both isomers and the fitted value should lie between 1.3 and 1.9~s. Using the same time fitting conditions as before we get $T_{1/2}$ = 1.70(3)~s. As discussed in subsection \ref{beta-n} below, this value is mainly due to the $3^{-}$ isomer half-life, which is the state predominantly populated in the $\beta$-n decay of $^{81}$Zn.

Transitions arising from the $\beta^-$ decay of $^{81}$Zn have been identified from their time spectra after proton impact, which is consistent with the $^{81}$Zn half-life of 0.32(5) s adopted in \cite{NDS81}. In our experiment, the $^{81}$Zn half-life has been measured using the time spectrum gated directly on three of the strongest $^{81}$Ga transitions of 351, 452 and 1341 keV (see Fig.~\ref{half_lives}a). A simple exponential decay plus constant background function has been used in the time range from 700 ms (with a slight delay after the end of implantation) to 2400 ms, restricting the time between proton
impact on target to two or more cycles (2.4 s or longer). 
The weighted mean value obtained yields $T_{1/2}$ = 290(4) ms, in agreement with the recent literature values \cite{pad10,xu14}.

\subsection{\texorpdfstring{$^{\bm{81}}$}{81}Z\lowercase{n} \texorpdfstring{$\bm{\beta^{-}}$}{Beta} decay to \texorpdfstring{$^{\bm{81}}$}{81}G\lowercase{a}}

\begin{figure}[ht!]
\centering
	\centerline{\includegraphics[width=\columnwidth]{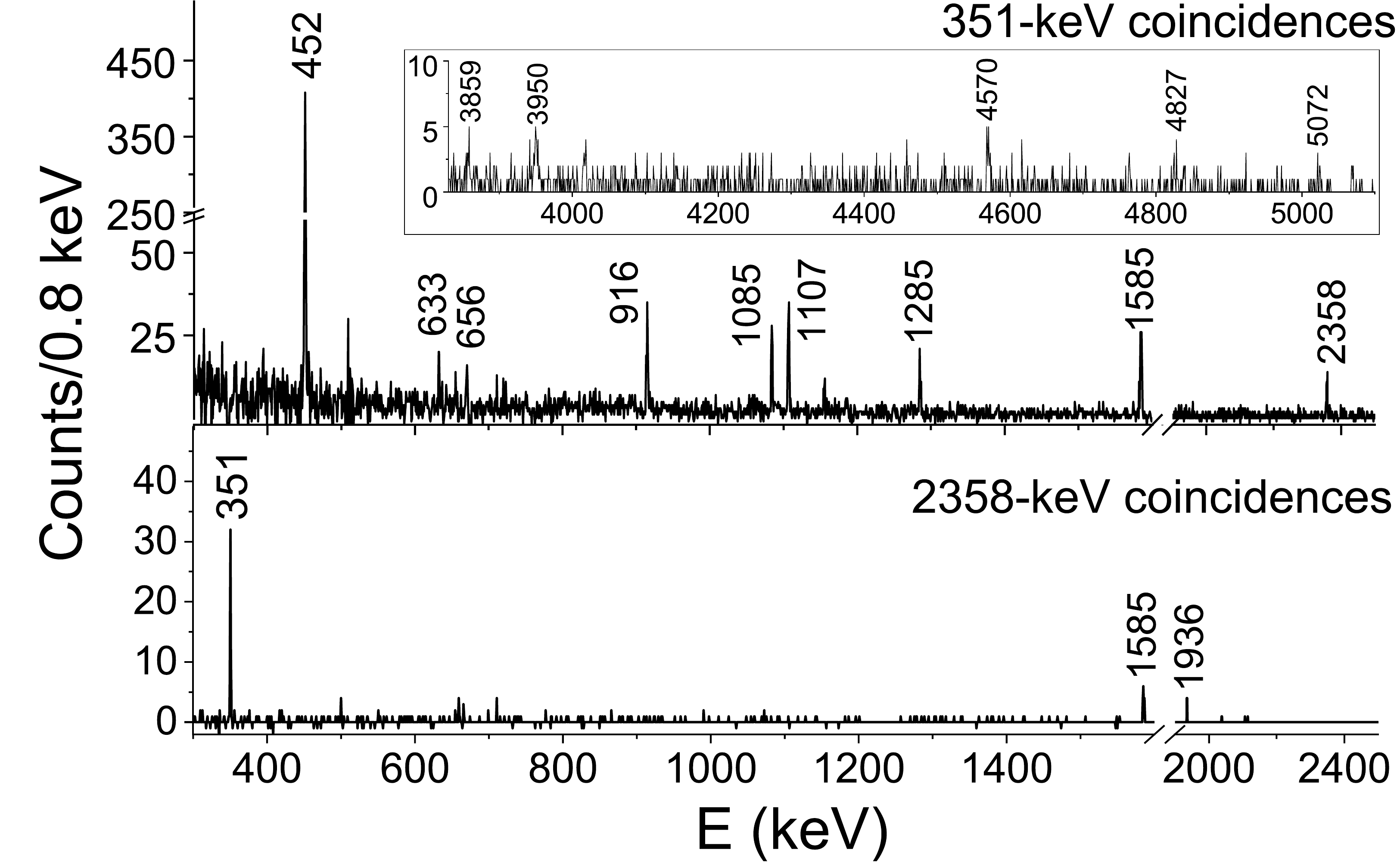}}
	\caption{\label{351_2358_coinc} $\gamma$-$\gamma$ coincidence spectra gated by the strongest, 351-keV transition (top) and the 2358-keV line (bottom).}
\end{figure}

The decay scheme of $^{81}$Ga has been extended using coincidences with previously known transitions employing the $\gamma$-$\gamma$ coincidence spectrum between both HPGe detectors. Figure~\ref{351_2358_coinc} shows the energy spectra in coincidence with the 351- and 2358-keV transitions. Note that $\gamma$ rays up to 5 MeV are registered in coincidence with the strong 351-keV $^{81}$Ga transition. 
Table~\ref{beta_gamma_subtracted} summarizes the information about the $\gamma$ transitions associated with the decay of $^{81}$Zn to $^{81}$Ga. The relative intensities of the $\gamma$-ray transitions were extracted using the full-energy peak areas from the $\beta$-gated $\gamma$-ray spectrum
and were normalized to the strongest transition at 351 keV. 
\newpage

\LTcapwidth=\textwidth
\begingroup
\begin{longtable*}{llllc}
\caption{\label{beta_gamma_subtracted} Gamma transitions in the decay of $^{81}$Zn to $^{81}$Ga. For those placed in the decay scheme, the initial and final level energies are given in the second and third columns. Relative intensities, normalized to 100 units for the 351-keV transition, are provided. The strongest transitions observed in $\gamma$-$\gamma$ coincidences are given in the last column.}\\	
\hline\hline
\textbf{$\bm{E_{\gamma}}$(keV)} & \textbf{$\bm{E_{ level}^{i}}$ (keV)} & \textbf{$\bm{E_{level}^{f}}$ (keV)} & \textbf{$\bm{I_{\gamma}^{rel}}$ $\bm{^a}$} & \textbf{Main $\bm{\gamma}$-$\bm{\gamma}$ coincidences}\\
\hline
\endfirsthead
\multicolumn{5}{c}%
{\tablename\ \thetable\ -- \textit{Continued}} \\
\hline
\textbf{$\bm{E_{\gamma}}$(keV)} & \textbf{$\bm{E_{ level}^{i}}$ (keV)} & \textbf{$\bm{E_{level}^{f}}$ (keV)} & \textbf{$\bm{I_{\gamma}^{rel}}$ $\bm{^a}$} & \textbf{Main $\bm{\gamma}$-$\bm{\gamma}$ coincidences}\\
\hline
\hline
\endhead
\hline \multicolumn{5}{c}{\textit{Continued on next page}} \\
\endfoot
\hline
\multicolumn{5}{l}{{\footnotesize $^a$~For absolute intensity per 100 parent decays, multiply by 0.374(22).}}\\
\multicolumn{5}{l}{{\footnotesize $^b$~Weak transition, not observed in $\gamma$-$\gamma$ coincidences. Tentatively placed in the level scheme.}}\\
\endlastfoot
\hline
			\textbf{333.3 \textit{2}} & 2285.6 \textit{1} & 1952.4 \textit{2} & 0.80 \textit{4} & {\footnotesize 611, 1341, 2009}\\
			\textbf{351.1 \textit{1}} & 351.1 \textit{1} & 0.0 & 100 \textit{4} & {\footnotesize 452, 633, 656, 916, 1085, 1107,} \\
			& & & &  {\footnotesize 1155, 1185, 1251, 1285, 1585,}\\
			& & & &  {\footnotesize 1847, 2065, 2358, 2807, 2838,}\\
			& & & &  {\footnotesize 3375, 3403, 3558, 3598, 3764,}\\
		  & & & &  {\footnotesize 3859, 3944, 3950, 4018, 4250,}\\
		  & & & &  {\footnotesize 4463, 4570, 4762, 4827, 4840,}\\
			& & & &  {\footnotesize 5024, 5072}\\
			\textbf{451.6 \textit{1}} & 802.5 \textit{1} & 351.1 \textit{1} & 20.1 \textit{7} & {\footnotesize 351, 633, 4619, 4857}\\
			\textbf{478.2$^b$ \textit{2}} & 1936.4 \textit{1} & 1458.3 \textit{1} & 0.35 \textit{3} & \\
			\textbf{611.4 \textit{1}} & 1952.4 \textit{2} & 1341.0 \textit{1} & 1.8 \textit{1} &  {\footnotesize 333, 1341}\\
			\textbf{632.9 \textit{1}} & 1435.5 \textit{1} & 802.5 \textit{1} &  0.90 \textit{7} & {\footnotesize 351, 452}\\
			\textbf{655.8$^b$ \textit{2}} & 1458.3 \textit{1} & 802.5 \textit{3} & 0.59 \textit{6} & {\footnotesize 351}\\
			\textbf{802.4 \textit{1}} & 802.5 \textit{3} & 0.0 & 4.9 \textit{2} & \\
			\textbf{884.8 \textit{2}} & 2285.6 \textit{1} & 1400.7 \textit{2} & 0.95 \textit{8} & {\footnotesize 1401, 2009}\\
			\textbf{894.1$^b$ \textit{1}} & 2830.7 \textit{3} & 1936.4 \textit{1} & 0.84 \textit{7}& \\
			\textbf{915.5 \textit{4}} & 1266.7 \textit{3} &  351.1 \textit{1} & 3.0 \textit{2} & {\footnotesize 351}\\
			\textbf{944.4 \textit{4}} & 2285.6 \textit{1} &  1341.0 \textit{1} & 1.33 \textit{8} & {\footnotesize 1341}\\
			\textbf{1084.7 \textit{5}} & 1435.5 \textit{1} & 351.1 \textit{1} & 3.0 \textit{2} & {\footnotesize 351}\\
			\textbf{1107.4 \textit{2}} & 1458.3 \textit{1} & 351.1 \textit{1} & 5.7 \textit{3} & {\footnotesize 351}\\
			\textbf{1155.0 \textit{2}} & 1506.3 \textit{1} & 351.1 \textit{1} & 0.68 \textit{8} & {\footnotesize 351}\\
			\textbf{1185.2$^b$ \textit{2}} & 5485.9 \textit{3} & 4301.6 \textit{4} & 0.8 \textit{1} & {\footnotesize 351}\\
			\textbf{1250.9 \textit{2}} & 2686.5 \textit{2} & 1435.5 \textit{1} & 0.58 \textit{7} & {\footnotesize 351, 1085}\\
			\textbf{1266.9 \textit{6}} & 1266.7 \textit{3} & 0.0 & 0.79 \textit{8} & \\
			\textbf{1285.3 \textit{1}} & 1636.4 \textit{2} & 351.1 \textit{2} & 2.7 \textit{2} & {\footnotesize 351}\\
			\textbf{1341.0 \textit{1}} & 1341.0 \textit{1} & 0.0 & 10.5 \textit{6} & {\footnotesize 333, 611, 944, 2009}\\
			\textbf{1400.7 \textit{1}} & 1400.7 \textit{2} & 0.0 & 11.0 \textit{6} & {\footnotesize 885}\\
			\textbf{1458.3 \textit{2}} & 1458.3 \textit{1} & 0.0 & 5.0 \textit{3} & \\
			\textbf{1506.4 \textit{1}} & 1506.3 \textit{1} & 0.0 & 8.5 \textit{5} & \\
			\textbf{1585.5 \textit{1}} & 1936.4 \textit{1} & 351.1 \textit{1} & 7.0 \textit{4} & {\footnotesize 351, 2358}\\
			\textbf{1847.2 \textit{4}} & 2198.3 \textit{4} & 351.1 \textit{1} & 0.9 \textit{1} & {\footnotesize 351}\\
			\textbf{1936.3 \textit{2}} & 1936.4 \textit{1} & 0.0 & 8.1 \textit{5} & {\footnotesize 2358}\\
			\textbf{2009.2 \textit{2}} & 4294.9 \textit{1} & 2285.6 \textit{1} & 4.8 \textit{3} & {\footnotesize 333, 611, 885, 1341, 2285}\\
			\textbf{2065.5 \textit{3}} & 2416.6 \textit{3} & 351.1 \textit{1} & 0.57 \textit{8} & {\footnotesize 351}\\
			\textbf{2285.5 \textit{2}} & 2285.6 \textit{7} & 0.0 & 3.7 \textit{2} & {\footnotesize 2009} \\
			\textbf{2358.4 \textit{2}} & 4294.9 \textit{1} & 1936.4 \textit{1} & 10.9 \textit{7} & {\footnotesize 351, 1585, 1936}\\
			\textbf{2686.6 \textit{4}} & 2686.5 \textit{2} & 0.0 & 1.8 \textit{2} & \\
			\textbf{2788.4 \textit{3}} & 2788.4 \textit{3} & 0.0 & 1.8 \textit{2} & \\
			\textbf{2807.0 \textit{3}} & 3158.1 \textit{4} & 351.1 \textit{1} & 1.0 \textit{1} & {\footnotesize 351}\\
			\textbf{2830.7 \textit{3}} & 2830.7 \textit{3} & 0.0 & 1.5 \textit{2} & \\
			\textbf{2838.2 \textit{7}} & 3189.3 \textit{7} & 351.1 \textit{1} & 1.0 \textit{1} & {\footnotesize 351}\\
			\textbf{3374.7 \textit{6}} & 3725.8 \textit{6} & 351.1 \textit{1} & 2.4 \textit{3} & {\footnotesize 351}\\
			\textbf{3402.7 \textit{4}} & 3753.8 \textit{4} & 351.1 \textit{1} & 2.1 \textit{3} & {\footnotesize 351}\\
			\textbf{3558.5 \textit{5}} & 3909.6 \textit{4} & 351.1 \textit{1} & 1.5 \textit{2} & {\footnotesize 351}\\
			\textbf{3598.2 \textit{5}} & 3949.3 \textit{5} & 351.1 \textit{1} & 2.1 \textit{3} & {\footnotesize 351}\\
			\textbf{3763.6 \textit{7}} & 4114.7 \textit{7} & 351.1 \textit{1} & 1.4 \textit{3} & {\footnotesize 351}\\
			\textbf{3858.5 \textit{4}} & 4209.3 \textit{3} & 351.1 \textit{1} & 4.1 \textit{6} & {\footnotesize 351}\\
			\textbf{3909.7 \textit{8}} & 3909.6 \textit{4} & 0.0 & 1.0 \textit{3} & \\
			\textbf{3943.9 \textit{5}} & 4294.9 \textit{1} & 351.1 \textit{1} & 2.2 \textit{4} & 351 \\
			\textbf{3950.5 \textit{4}} & 4301.6 \textit{4} & 351.1 \textit{1} & 2.8 \textit{4} & {\footnotesize 351}\\
			\textbf{4017.8 \textit{5}} & 4369.0 \textit{4} & 351.1 \textit{1} & 4.2 \textit{6} & {\footnotesize 351}\\
			\textbf{4208.5 \textit{6}} & 4209.3 \textit{3} & 0.0 & 1.1 \textit{2} & \\
			\textbf{4250.5 \textit{5}} & 4601.6 \textit{5} & 351.1 \textit{1} & 0.5 \textit{2} & {\footnotesize 351}\\
			\textbf{4295.4 \textit{4}} & 4294.9 \textit{1} & 0.0 & 4.6 \textit{6} & \\
			\textbf{4328.9 \textit{6}} & 5131.4 \textit{4} & 802.5 \textit{1} & 1.0 \textit{3} & {\footnotesize 351, 452}\\
			\textbf{4369.2 \textit{4}} & 4369.0 \textit{4} & 0.0 & 2.9 \textit{6} & \\
			\textbf{4374.6 \textit{7}} & 5177.8 \textit{3} & 802.5 \textit{1} & 1.3 \textit{3} &  {\footnotesize 351, 452}\\
			\textbf{4463.2 \textit{6}} & 4814.3 \textit{8} & 351.1 \textit{1} & 1.9 \textit{3} & {\footnotesize 351}\\
			\textbf{4570.0 \textit{4}} & 4921.1 \textit{4} & 351.1 \textit{1} & 6.8 \textit{8} &  {\footnotesize 351}\\
			\textbf{4618.9 \textit{7}} & 5421.9 \textit{2} & 802.5 \textit{1} & 1.7 \textit{3} &  {\footnotesize 351, 452}\\
			\textbf{4761.9 \textit{10}} & 5113.2 \textit{8} & 351.1 \textit{1} & 1.9 \textit{3} &  {\footnotesize 351}\\
			\textbf{4826.9 \textit{4}} & 5177.8 \textit{3} & 351.1 \textit{1} & 2.4 \textit{3} &  {\footnotesize 351}\\
			\textbf{4839.8 \textit{7}} & 5190.9 \textit{7} & 351.1 \textit{1} & 2.9 \textit{4} &  {\footnotesize 351}\\
			\textbf{4856.6 \textit{5}} & 5658.6 \textit{3} & 802.5 \textit{1} & 2.7 \textit{3} & {\footnotesize 351, 452}\\
			\textbf{4880.4 \textit{4}} & 4880.4 \textit{4} & 0.0 & 7.2 \textit{9} & \\
			\textbf{5024.0 \textit{5}} & 5375.1 \textit{5} & 351.1 \textit{1} & 0.5 \textit{1} & {\footnotesize 351} \\
			\textbf{5072.0 \textit{5}} & 5421.9 \textit{2} & 351.1 \textit{1} & 1.1 \textit{2} &  {\footnotesize 351}\\
			\textbf{5113.6 \textit{6}} & 5113.2 \textit{8} & 0.0 & 1.3 \textit{3} & \\
			\textbf{5178.2 \textit{5}} & 5177.8 \textit{3} & 0.0 & 4.5 \textit{6} & \\
			\textbf{5421.6 \textit{3}} & 5421.9 \textit{2} & 0.0 & 0.8 \textit{2} & \\
			\textbf{5475.5 \textit{5}} & 5475.5 \textit{5} & 0.0 & 2.5 \textit{4} & \\
			\textbf{5485.1 \textit{5}} & 5485.9 \textit{3} & 0.0 & 4.6 \textit{6} & \\
			\textbf{5657.4 \textit{5}} & 5658.6 \textit{3} & 0.0 & 0.7 \textit{2} & \\
			\textbf{5694.8 \textit{7}} & 5695.5 \textit{7} & 0.0 & 0.7 \textit{2} & \\
			\textbf{5726.9 \textit{4}} & 5726.9 \textit{4} & 0.0 & 0.6 \textit{1} & \\
			\textbf{5831.0 \textit{5}} & 5831.0 \textit{5} & 0.0 & 4.4 \textit{6} & \\
			\textbf{5863.5 \textit{3}} & 5863.5 \textit{3} & 0.0 & 0.6 \textit{1} & \\
			\textbf{5903.9 \textit{8}} & 5903.9 \textit{8} & 0.0 & 2.0 \textit{3} & \\
			\textbf{5936.1 \textit{6}} & 5936.1 \textit{6} & 0.0 & 0.4 \textit{1} & \\
			\textbf{5969.2 \textit{7}} & 5969.2 \textit{7} & 0.0 & 2.3 \textit{3} & \\
			\textbf{6150.5 \textit{7}} & 6150.5 \textit{7} & 0.0 & 1.8 \textit{3} & \\
			\textbf{6212.8 \textit{4}} & 6212.8 \textit{4} & 0.0 & 0.29 \textit{6} & \\
			\textbf{6236.1 \textit{5}} & 6236.1 \textit{5} & 0.0 & 0.24 \textit{6} & \\
			\textbf{6295.3 \textit{5}} & 6295.3 \textit{5} & 0.0 & 0.26 \textit{6} & \\
			\textbf{6405.2 \textit{5}} & 6405.2 \textit{5} & 0.0 & 0.27 \textit{6} & \\
			\textbf{6434.6 \textit{4}} & 6434.6 \textit{4} & 0.0 & 0.09 \textit{3} & \\
\hline\hline
\end{longtable*}
\endgroup

\begin{figure*}[ht!]
	\centerline{\includegraphics[scale=0.85, angle=90]{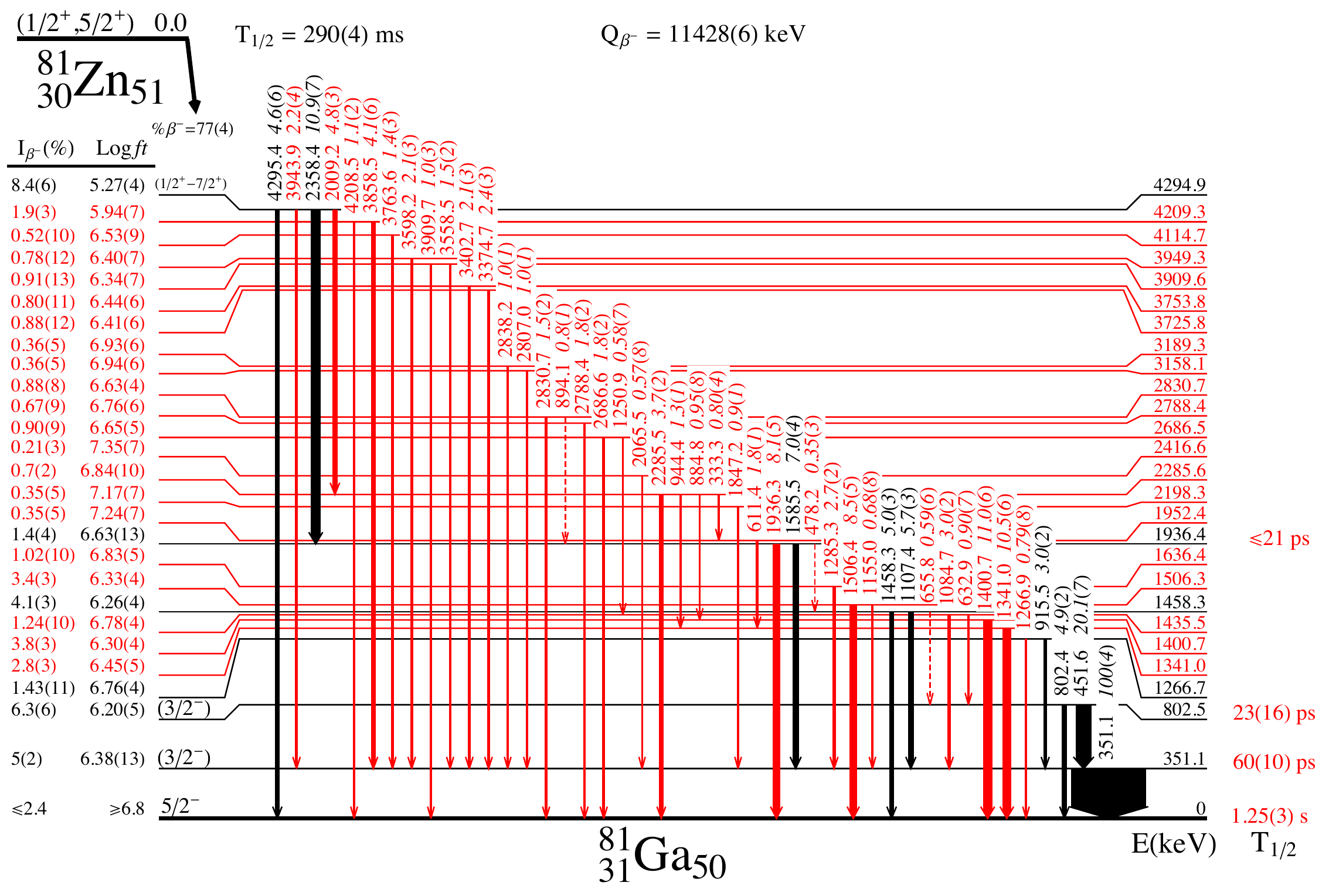}}
	\caption[width=\textwidth]{\label{LevelScheme_81Ga_low} Level scheme of $^{81}$Ga up to 4.3~MeV in energy populated following the $\beta$ decay of $^{81}$Zn. Dashed arrows indicate tentatively placed transitions. For the sake of clarity the decay scheme has been split in two sections.}
\end{figure*}

\begin{figure*}[ht!]
	\centerline{\includegraphics[scale=0.69, angle=90]{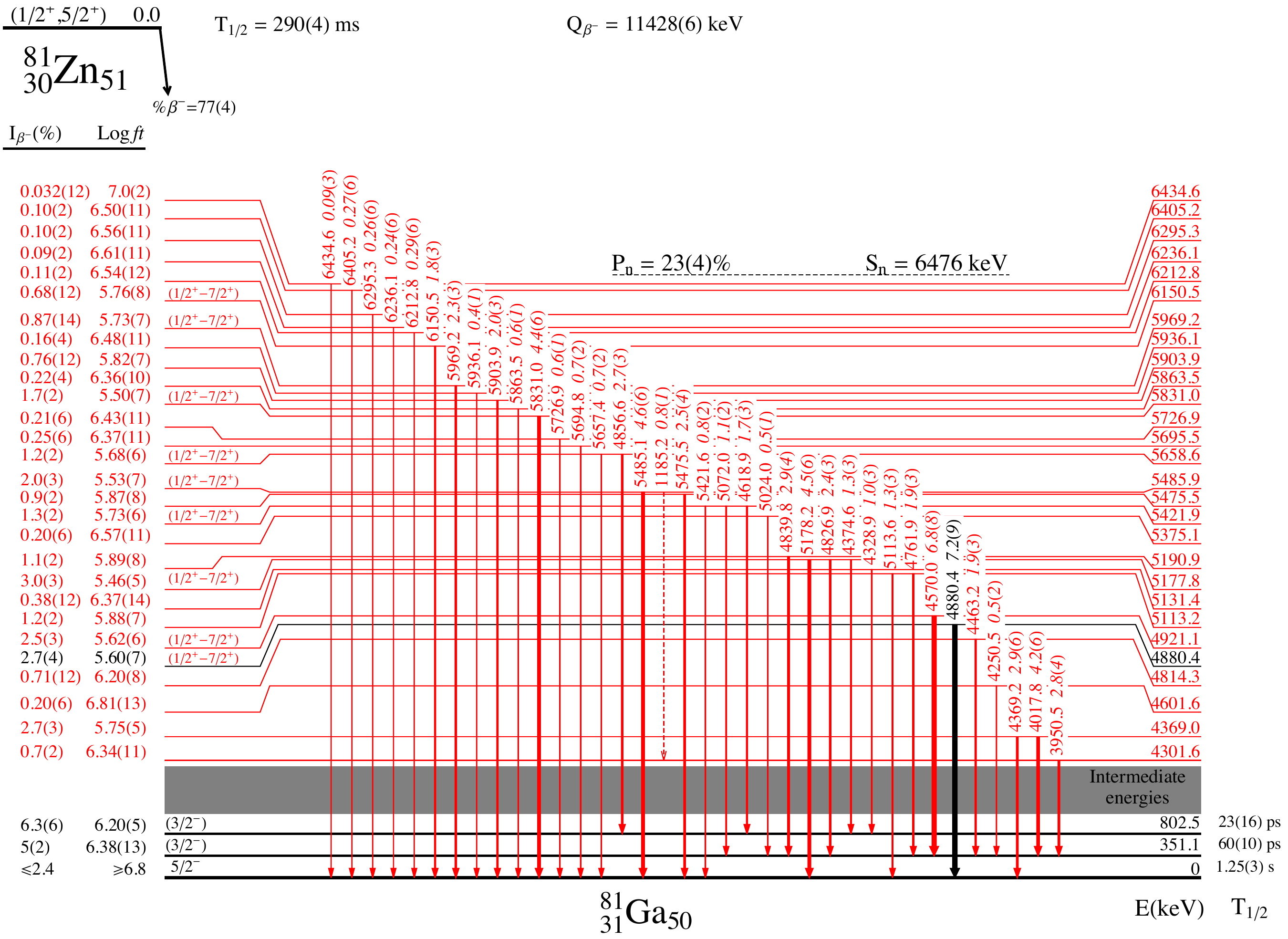}}
	\caption[width=\textwidth]{\label{LevelScheme_81Ga_high} Level scheme of $^{81}$Ga populated in the $\beta$ decay of $^{81}$Zn, containing the high-lying states between 4.3 and 6.5 MeV in energy. Dashed arrows indicate tentatively placed transitions.}
\end{figure*}

Based on the $\gamma$-$\gamma$ coincidences, 70 transitions that were not previously observed in Ref. \cite{pad10} have been placed in the level scheme, which is shown in Figs.~\ref{LevelScheme_81Ga_low}~and~\ref{LevelScheme_81Ga_high}. Weak transitions that were not observed in coincidence with strong $^{81}$Ga $\gamma$ rays have not been included, since they could also belong to the level scheme of $^{80}$Ga populated in the $\beta^-$n decay of $^{81}$Zn (see Sec.~\ref{betan}). Such is the case of 279-, 505-, 627-, 779-, 2627- and 2943-keV $\gamma$ rays. Their combined intensities amount to 1.1\% of the total $\gamma$ intensity. However, some of the weak $\gamma$ rays of 478, 656, 894 and 1185 keV, fit the energy differences between already established levels and were tentatively placed in the level scheme. They are marked with broken lines. The high-energy $\gamma$ rays not observed in coincidence with those at 351 and 452 keV were placed as de-exciting a state with the same energy. We note that the available energy window for $\beta^{-}$ decay is $Q_{\beta^-}=11428(6)$ keV \cite{AME2016}, compared to a value of $Q_{\beta^{-}n}=4953(6)$ keV {\cite{AME2016}} for $\beta$-delayed neutron emission. Therefore, $\gamma$ rays with energies above 5 MeV that follow the $^{81}$Zn half-life must belong to $^{81}$Ga and not to $^{80}$Ga.  

In this way, 47 excited states of $^{81}$Ga in the energy range up to the neutron separation energy of 6476(4)~keV {\cite{AME2016}} have been observed, 40 of them for the first time. We confirm the existence of 351.1-, 802.5-, 1266.7-, 1458.3-, 1936.4-, 4294.9- and 4880.4-keV levels, already seen in the latest $\beta$-decay study \cite{pad10}. The states identified as $(9/2^{-})$ and $(11/2^{-})$ in fission $\gamma$-ray spectroscopy \cite{dudouet19} are also observed at 1341.0 and 1952.4~keV \cite{paziy16PhD}.

\subsection{Beta-delayed neutron emission probability of $^{81}$Zn}
\label{beta-n}

To obtain the $\beta$-delayed neutron emission probability of $^{81}$Zn we compared the number of decays arising from the direct $^{81}$Zn $\beta$-decay chain, using the absolute intensities of the two strongest lines in $^{81}$Ge, at 216 and 828 keV \cite{NDS81}, to the $^{81}$Zn $\beta^{-}$n decay branch of the $A=80$ chain, taking the absolute intensities per 100 parent decays of 666-, 1207-, and 1645-keV lines from the $\beta$ decay of $^{80}$As to $^{80}$Se \cite{mcil71,kratz75}. We employ the literature value of 11.9(7)\% for the $^{81}$Ga $\beta^{-}$n branch \cite{NDS81} and apply a small correction factor coming from the $^{80}$Ga $\beta^{-}$n decay probability of 0.86(7)\%, also taken from the literature \cite{NDS79}. Determining the areas of the above mentioned transitions directly from the $\beta$-gated singles spectrum and taking into account the absolute intensities we obtain $P_n$=23(4)\% for $^{81}$Zn.

\subsection{Direct \texorpdfstring{$\bm{\beta}$}{Beta} feeding to the $^{81}$Ga ground state}

For the absolute $\beta$ feeding to be derived, it is necessary to obtain the ground-state (g.s.) $\beta$ feeding. Since there is no isomeric state reported for $^{81}$Ga, the total g.s. feeding, both $\gamma$ and $\beta$, proceeds through the $^{81}$Ga ground-state $\beta$ decay to states in $^{81}$Ge, and via the $\beta$-delayed neutron emission branch to states in $^{80}$Ge. A $\beta$-decaying isomer exists in $^{81}$Ge at 679 keV \cite{hoff81}, for which no $\gamma$-ray branch was observed. Therefore, these two states need to be considered in the $\beta$ decay of $^{81}$Ga, both for $\gamma$ and $\beta$ feeding. For the $\beta$-n branch from $^{81}$Ga we take an adopted $P_n$ value of 11.9(7)\% from \cite{NDS81}. In addition, for the $^{81}$Zn the $P_n$ value of 23(4)\% from our data is used, as described in subsection \ref{beta-n} above.

The $\gamma$-ray intensities in $^{81}$Ga and $^{81}$Ge are obtained from our data without time conditions, thus containing the short-lived and long-lived decay products from $^{81}$Zn and its daughters, and normalized to the strongest 351-keV transition in $^{81}$Ga, Table~\ref{beta_gamma_subtracted}. The total $\gamma$ intensity feeding the ground state of $^{81}$Ga is measured to be $I^{Ga}_{\gamma,gs}=203(4)$. In the decay of $^{81}$Ga, the $\gamma$-ray intensity that feeds directly the 679-keV isomer state and the ground state amounts to 101(3) and 95(3) in the same units, respectively, and thus the $\gamma$-ray feeding both states is 196(4) units.

To estimate the $\beta$-feeding intensity to the g.s. and 679-keV isomer in $^{81}$Ge we make use of the spin assignments of 9/2$^{+}$ and 1/2$^+$ \cite{hoff81}. These levels are therefore $\beta$-fed from the $^{81}$Ga 5/2$^-$ ground state via first-forbidden unique $\beta$ transitions with $\Delta{J=2}$, $\Delta{\pi}=yes$.
It is then reasonable to consider a lower limit of log$^{1U}{ft}$ = 8.5 (see Fig.~1 of {\cite{logft_review}}) for both states. The $\beta$ feeding calculated with these assumptions gives upper limits of 11.3\% for the 9/2$^{+}$ g.s. and 6.6\% for the 679-keV isomer represented in absolute units (5.7(56)\% and 3.3(33)\% were used for calculations). 

With these assumptions the value of the ground-state $\beta$ feeding in $^{81}$Ga is extracted from the intensity balance and is given with an upper limit of 2.4\%. This leads to log${ft}\geq{6.8}$, in good agreement with the systematics and selection rules for the first-forbidden non-unique $\beta$ decay transitions in the region. 

Using this value, the apparent $\beta$ feeding of the remaining levels, $I_{\beta}(E)$, is obtained by the intensity balance between feeding and de-exciting $\gamma$ rays. Internal conversion is neglected. High-energy transitions could have been missed or misplaced if coincidences are not observed, which means that the $\beta$ feeding would be slightly modified. We note the small energy  gap between the highest level at 6434.6 keV and the neutron separation energy, $S_n=6476(4)$ keV, which is still far from the available $\beta$-decay window, $Q_{\beta^-}=11428(6)$ keV {\cite{AME2016}}. With the $^{81}$Zn $\beta$-delayed neutron emission probability $P_{n}=23(4)\%$ and the ground-state feeding (taking $I^{Ga}_{\beta,gs}$= $1.2(12)\%$), an absolute normalization factor of 0.374(22) is obtained for the $\gamma$ intensities in the decay of $^{81}$Zn to $^{81}$Ga from the relative ones tabulated in Table~\ref{beta_gamma_subtracted}.
 
\subsection{\texorpdfstring{$^{\bm{81}}$}{81}Z\lowercase{n} \texorpdfstring{$\bm{\beta^{-}}$}{Beta}\lowercase{n} decay to \texorpdfstring{$^{\bm{80}}$}{81}G\lowercase{a}}
\label{betan}

\begin{figure}
	\centerline{\includegraphics[width=\columnwidth]{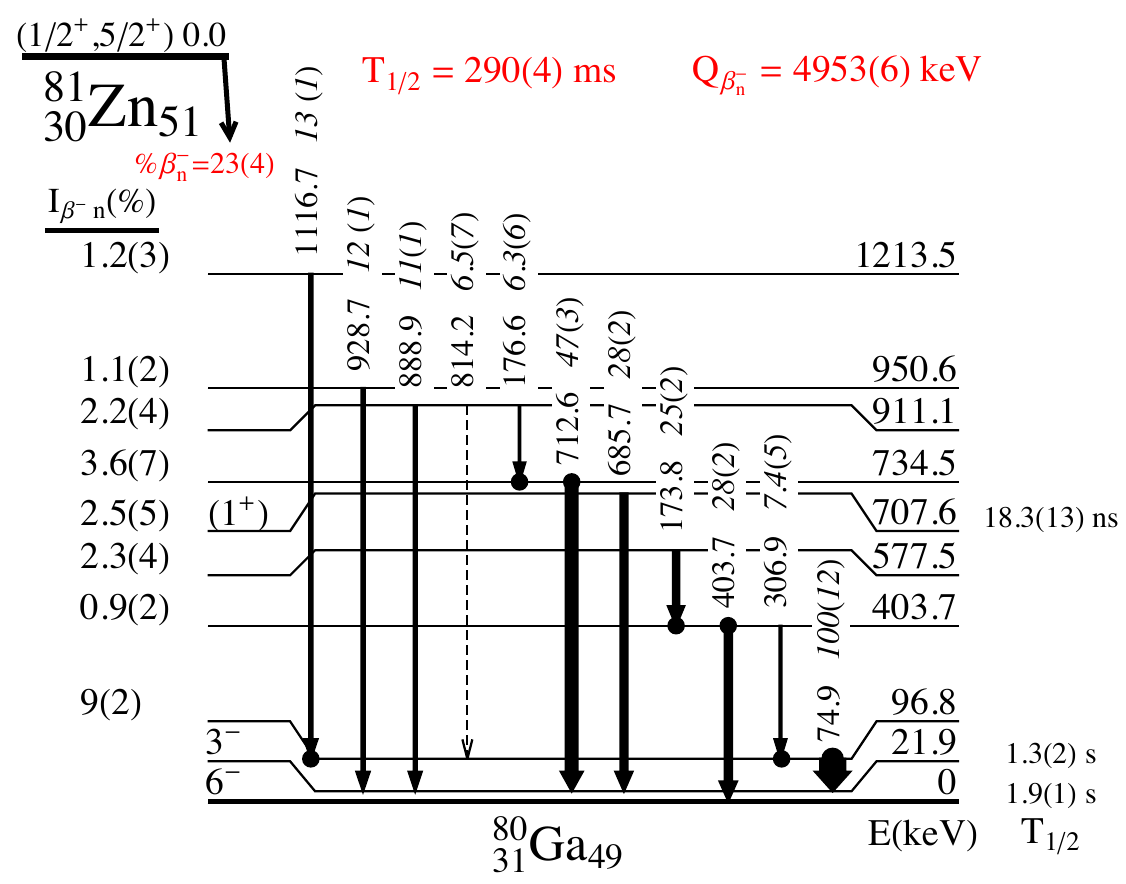}}
	\caption{\label{level_scheme_80Ga}Levels in $^{80}$Ga populated in the $\beta^-$n decay of $^{81}$Zn from our work. The half-lives of the 21.9-keV and ground state were previously reported in \cite{ver13}.}
\end{figure}

As discussed above, $\beta$-delayed neutron emission is energetically allowed for the decay of $^{81}$Zn, with a $^{81}$Ga neutron separation energy S$_n$= 6476(4) keV {\cite{AME2016}}, well within the $Q_{\beta^-}$ window. The analysis of the $\beta$-gated $\gamma$ spectrum has allowed 11 $\gamma$ transitions to be assigned to $^{80}$Ga populated following the $\beta^-$n decay of $^{81}$Zn. The nuclide $^{80}$Ga was studied at ISOLDE during the same experimental run, populated in the $\beta^-$ decay of $^{80}$Zn, and the results of the analysis were published by Lic\v{a} \textit{et al.} \cite{RAZ14}. The $\gamma$-$\gamma$ coincidence analysis provides information to place the observed $^{80}$Ga transitions de-exciting nine previously known low-energy levels. Our new scheme of $^{80}$Ga from $\beta^-$n decay of $^{81}$Zn plotted in Fig.~\ref{level_scheme_80Ga} is consistent with the structure from {\cite{RAZ14}}. 
Table~\ref{80Ga} contains the detailed information about the $\gamma$ transitions and fed energy levels. We neglect any direct feeding of the $6^-$ ground and the $3^-$ first isomeric states in $^{80}$Ga from the ($1/2^{+}$, $5/2^{+}$) ground state of $^{81}$Zn.
The apparent $\beta$-n feeding, $I_{\beta n}(E)$, is obtained from the intensity balance. Internal conversion is included, specifically for the
75-keV transition, by taking coefficients from \cite{bricc} and assuming dipole transitions.
It is worth noting that most of the population from the $\beta^-$n decay of $^{81}$Zn proceeds to the $3^-$ state at 22 keV.
The second isomer, with spin-parity $1^{+}$, is confirmed at 708 keV. We measured its half-life to be $T_{1/2}=18.3(13)$ ns using triple coincidences between the $\beta$ and two HPGe detectors. Our half-life for the 708-keV state has slightly less precision but is in perfect agreement with the value determined in \cite{RAZ14}.

\begin{table} [h]
\caption{\label{80Ga}Gamma transitions in $^{80}$Ga populated in the $\beta^-$-n decay of $^{81}$Zn. Intensities relative to the 74.9-keV transition, placements in the level scheme, and main $\gamma$-$\gamma$ coincidences are listed where available.}
\centering
\begin{tabular}{llllc}
\hline\hline
\textbf{$E\bm{_{\gamma}}$} & \textbf{$E\bm{_{level}^{i}}$} & \textbf{$E\bm{_{level}^{f}}$} & \textbf{$\bm{I_{\gamma}}$} & \\
(keV) & (keV) & (keV) &  (\%) & $\bm{\gamma}$-$\bm{\gamma}$\\
\hline
\textbf{74.9 \textit{1}} & 96.8 \textit{3} & 21.9 \textit{3} & 100 \textit{12} & {\footnotesize 307, 1117}\\
\textbf{173.8 \textit{4}} & 577.5 \textit{2} & 403.7 \textit{2} & 25 \textit{2} & {\footnotesize 404}\\
\textbf{176.6 \textit{1}} & 911.1 \textit{4} & 734.5 \textit{3} & 6.3 \textit{6} & {\footnotesize 713}\\
\textbf{306.9 \textit{2}} & 403.7 \textit{2} & 96.8 \textit{3} & 7.4 \textit{5} & {\footnotesize 75}\\
\textbf{403.7 \textit{2}} & 403.7 \textit{2} & 0.0 & 28 \textit{2} & {\footnotesize 174}\\
\textbf{685.7 \textit{1}} & 707.6 \textit{3} & 21.9 \textit{3} & 28 \textit{2} & \\
\textbf{712.6 \textit{6}} & 734.5 \textit{3} & 21.9 \textit{3} & 47 \textit{3} & {\footnotesize 177}\\
\textbf{814.2$^{a}$ \textit{2}} & 911.1 \textit{4} & 96.8 \textit{3} & 6.5 \textit{7} & \\
\textbf{888.9 \textit{3}} & 911.1 \textit{4} & 21.9 \textit{3} & 11 \textit{1} & \\
\textbf{928.7 \textit{5}} & 950.6 \textit{6} & 21.9 \textit{3} & 12 \textit{1} & \\
\textbf{1116.7 \textit{3}} & 1213.5 \textit{4} & 96.8 \textit{3} & 13 \textit{1} & {\footnotesize 75}\\
\hline
\multicolumn{5}{l}{{\footnotesize $^a$~Weak transition, not observed in $\gamma$-$\gamma$ coincidences.}}\\
\multicolumn{5}{l}{{\footnotesize Tentatively placed in the level scheme.}}\\
\end{tabular}
\end{table}

\subsection{Half-lives of the excited states of \texorpdfstring{$^{\bm{81}}$}{81}G\lowercase{a}}

\begin{figure}
\includegraphics[scale=0.43]{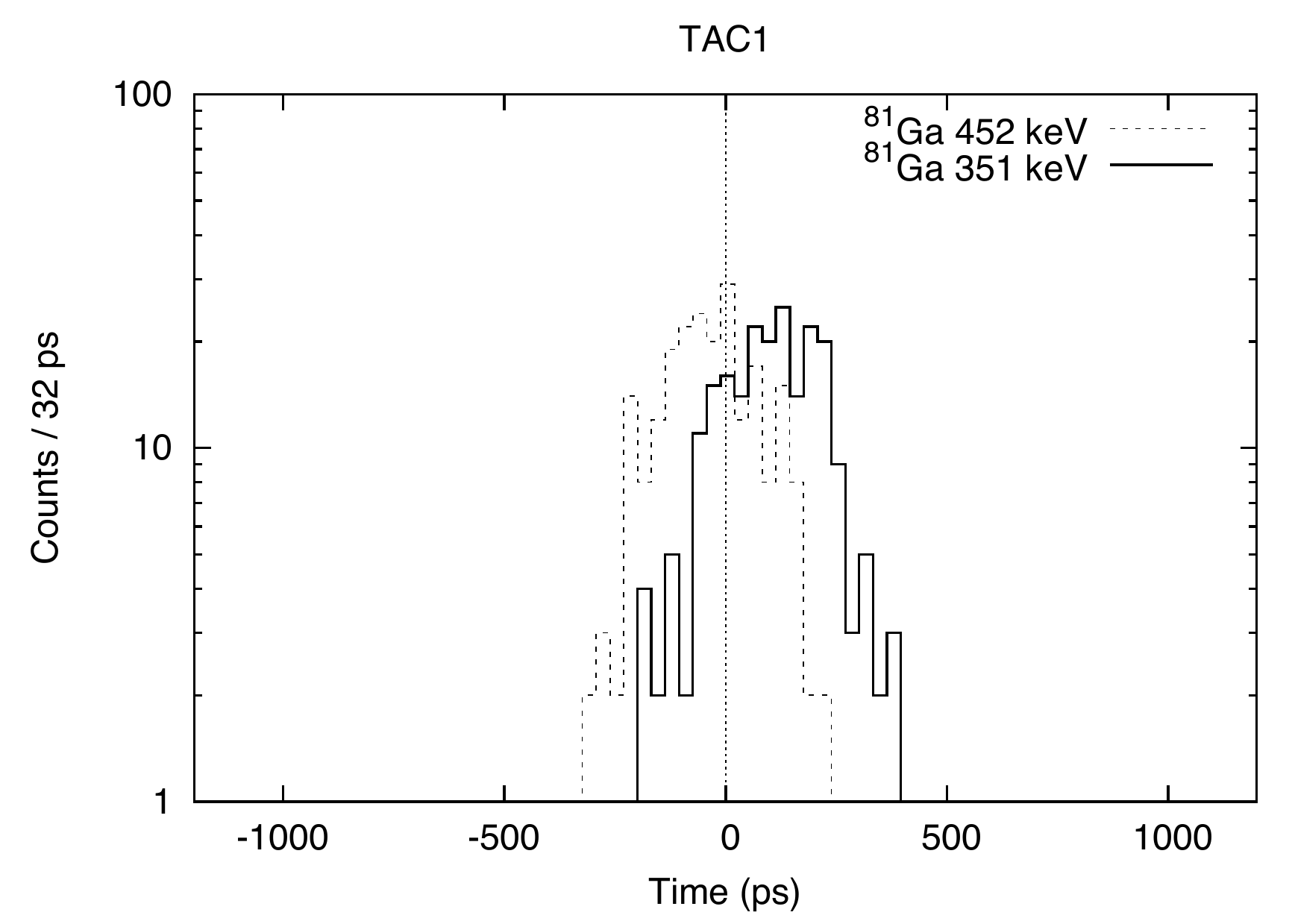}
\includegraphics[scale=0.43]{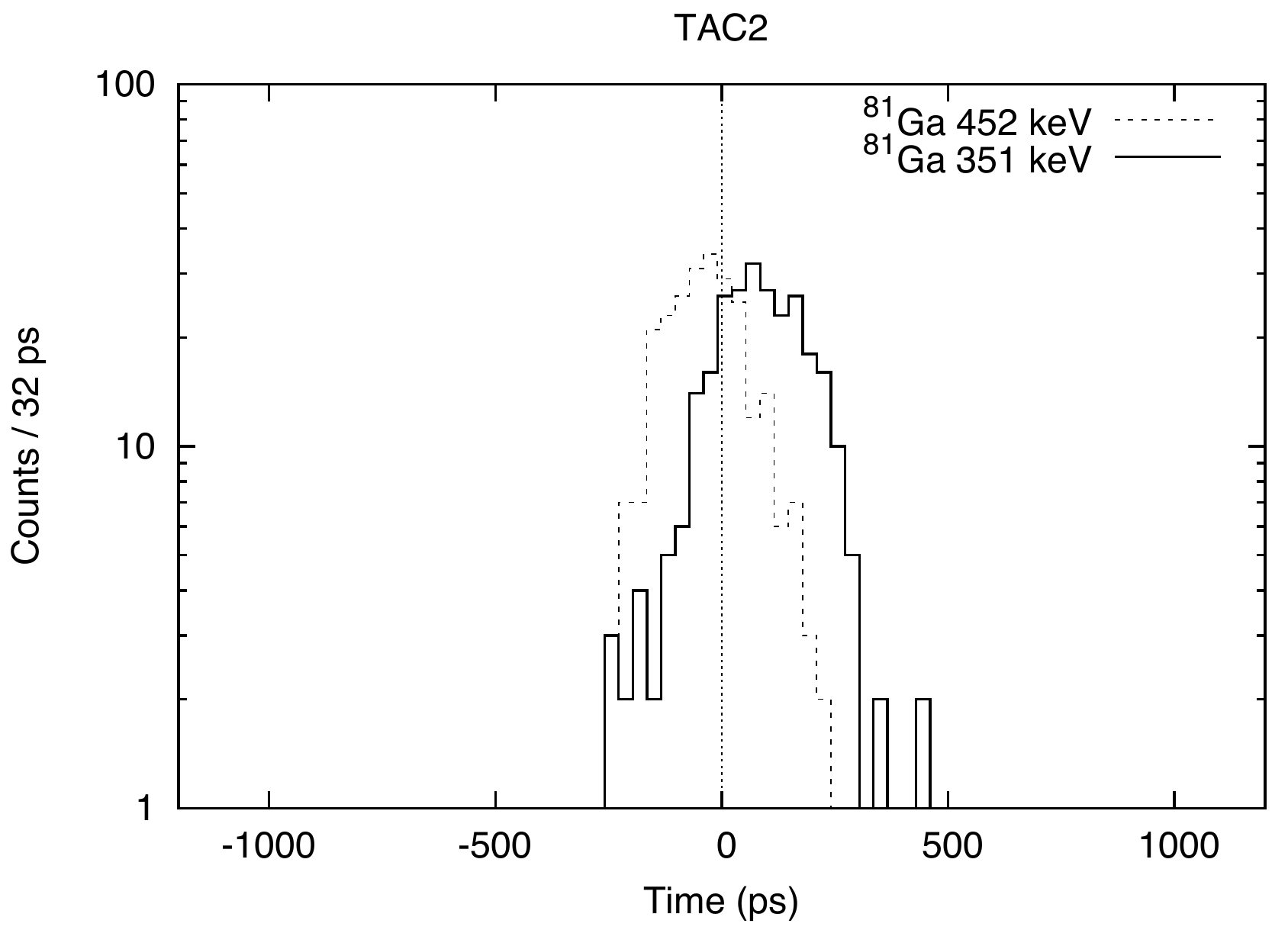}
	\caption{\label{CS_351keV} Time spectra obtained in triple $\beta\gamma\gamma$(t) coincidences with the
351-keV $\gamma$ transition selected in the HPGe detectors and the 452-keV in the LaBr$_{3}$(Ce) detector (dotted line) and 
with reversed $\gamma$ gates (solid line). The left panel shows the TAC spectra for the first LaBr$_{3}$(Ce) and the right panel
the spectra for the second LaBr$_{3}$(Ce). The time distributions do not include timing corrections of the prompt positions and Compton background contributions. Once corrected for these, the difference of the centroid positions of the time distributions yields the mean-life of the 351-keV level. 
See text for details.}
\end{figure}

Two strong sequential transitions of 351 and 452 keV are observed in the level scheme of $^{81}$Ga (Fig.~\ref{LevelScheme_81Ga_low} and Fig.~\ref{LevelScheme_81Ga_high}). The first one de-excites the first excited state of the same energy while the second one comes from the 802-keV energy level. Selecting the 351-keV transition in the HPGe detector and the 452-keV one in the LaBr$_{3}$(Ce) detectors, the $\beta$-LaBr$_{3}$(Ce)
time difference distribution is due to the lifetime of the 802-keV level plus the contributions from the lifetimes of higher lying levels. By reversing the gates, selecting the 452-keV line now in the HPGe and the 351-keV one in the LaBr$_{3}$(Ce) detector, the observed time delayed spectrum arises from the lifetime of both the 802-keV and the 351-keV levels, plus the contributions from higher lying states. The difference between the centroids of both time distributions, once corrected for the different prompt positions at 351 and 452 keV (using the FEP response curve and their Compton background contribution), yields the mean-life of the 351-keV level. Figure~\ref{CS_351keV} shows two plots that illustrate the time distributions under these conditions. The time difference between their centroids shown in the figure is not yet corrected by the effect of prompt position and the Compton background response. After corrections, the centroid shift method gives the values of $\tau = 92(15)$ ps for the first LaBr$_{3}$(Ce) detector and 80(13) ps for the second one. We take the average of both values and uncertainties, which leads to a $T_{1/2}=60(10)$ ps half-life.

\begin{table*}
\centering
\caption{Summary of half-lives of excited states in $^{81}$Ga, and experimental $B(M1)$ and $B(E2)$ reduced transition probabilities for the de-exciting transitions, assuming pure multipolarities. They are compared to the theoretical values calculated with the JUN45 and jj44b effective interactions (see text for details).}
\label{transition_prob}
\begin{tabular}{cccccccccc}
\hline
\textbf{$\bm{E_{level}}$} & \textbf{$\bm{T_{1/2}}$} & $\bm{J^{\pi}}$ & \textbf{$\bm{E_{\gamma}$}} & \multicolumn{3}{c}{\textbf{$\bm{B(M1)}$ W.u.}} & \multicolumn{3}{c}{\textbf{$\bm{B(E2)}$ W.u.}} \\ 
\textbf{(keV)} & \textbf{(ps)} & & \textbf{(keV)} & \textbf{EXP} & \textbf{JUN45} & \textbf{jj44b} & \textbf{EXP} & \textbf{JUN45} & \textbf{jj44b} \\
\hline
351  & 60(10)     & ($3/2^{-}$) & 351  & 8.5(14)$\times{10^{-3}}$   & 5.0$\times{10^{-4}}$ & 1.1$\times{10^{-4}}$ & 85(14)                  & 2.9  & 3.2 \\
802  & 23(16)     & ($3/2^{-}$) & 452  & 8(6)$\times{10^{-3}}$      & 0.23   & 0.06      & 50(30)   & 2.4  & 0.7 \\
     &            &             & 802  & 4(2)$\times{10^{-4}}$      & 9.1$\times{10^{-4}}$  & 2.1$\times{10^{-4}}$      & 0.7(5)       & 6.9 & 9.7 \\ 
1936 & $\leq{21}$ &             & (478) & $\geq$ 2.0$\times{10^{-4}}$ &       &           & $\geq{1.1}$ &      & \\
 &  &             & 1585 & $\geq$ 1.1$\times{10^{-4}}$ &       &           & $\geq{0.056}$ &      & \\
     &            &             & 1936 & $\geq$ 7.1$\times{10^{-5}}$ &       &           & $\geq{0.024}$ &      & \\
\hline
\end{tabular}
\end{table*}

As a cross-check we have tried to de-convolute the slope in the time spectra in $\beta\gamma$(t) and $\beta\gamma\gamma$(t) coincidences by selecting the 351-keV transition in the LaBr$_{3}$(Ce) detectors and fixing the prompt distribution to that given by the 452-keV transition. Although the result is limited by statistics, it is consistent with a slope that yields a half-life of the order of 50~ps. 

A similar procedure to that described above for the 351-keV level is applied to measure the lifetime of the 1936-keV state, using in this case the coincident 1585- and 2358-keV transitions. The results are at the limit of sensitivity and yield $\tau$ = 20(18) ps and $\tau$ = 6(16) ps respectively. We take the average value of $\tau=13(17)$ ps resulting in a one-sigma upper limit of $T_{1/2}\leq{21}$ ps for this level. 

The half-life of the second excited state at 802 keV is measured by absolute comparison using parallel transitions \cite{mach91}. 
The high-lying states in $^{81}$Ga are characterized by short half-lives below $\sim$1 ps. 
Several high-energy $\gamma$ transitions are in coincidence with the 351-keV $\gamma$ ray (Fig.~\ref{LevelScheme_81Ga_low} and Fig.~\ref{LevelScheme_81Ga_high}). By selecting those in the HPGe detectors and the 351-keV one in the LaBr$_{3}$(Ce) detector, 
the $\beta$-LaBr$_{3}$(Ce) time difference will arise from the 351-keV state lifetime. This can be compared with the time distribution resulting from the selection of the 452-keV transition in the HPGe and the 351-keV one in the LaBr$_{3}$(Ce) detectors, which is due to both the 351-keV and 802-keV lifetimes. The difference between centroid positions, once corrected by the calibrations, gives an average of $\tau=34(22)$ ps or $T_{1/2}=23(16)$ ps for the 802-keV level half-life.

The previously unknown half-lives obtained from this measurement are summarized in Tab.~\ref{transition_prob}. Using the lifetimes and $\gamma$-ray branching from our level scheme, the transition probabilities for the de-exciting lines have been calculated for the most probable multipolarities. The theoretical evaluation of conversion coefficients \cite{bricc} for these transitions show that all of them are well below 1\% and thus were neglected. Pure transitions are assumed for the experimental values. 

\begin{figure*}[ht!]
	\centerline{\includegraphics[scale=0.5]{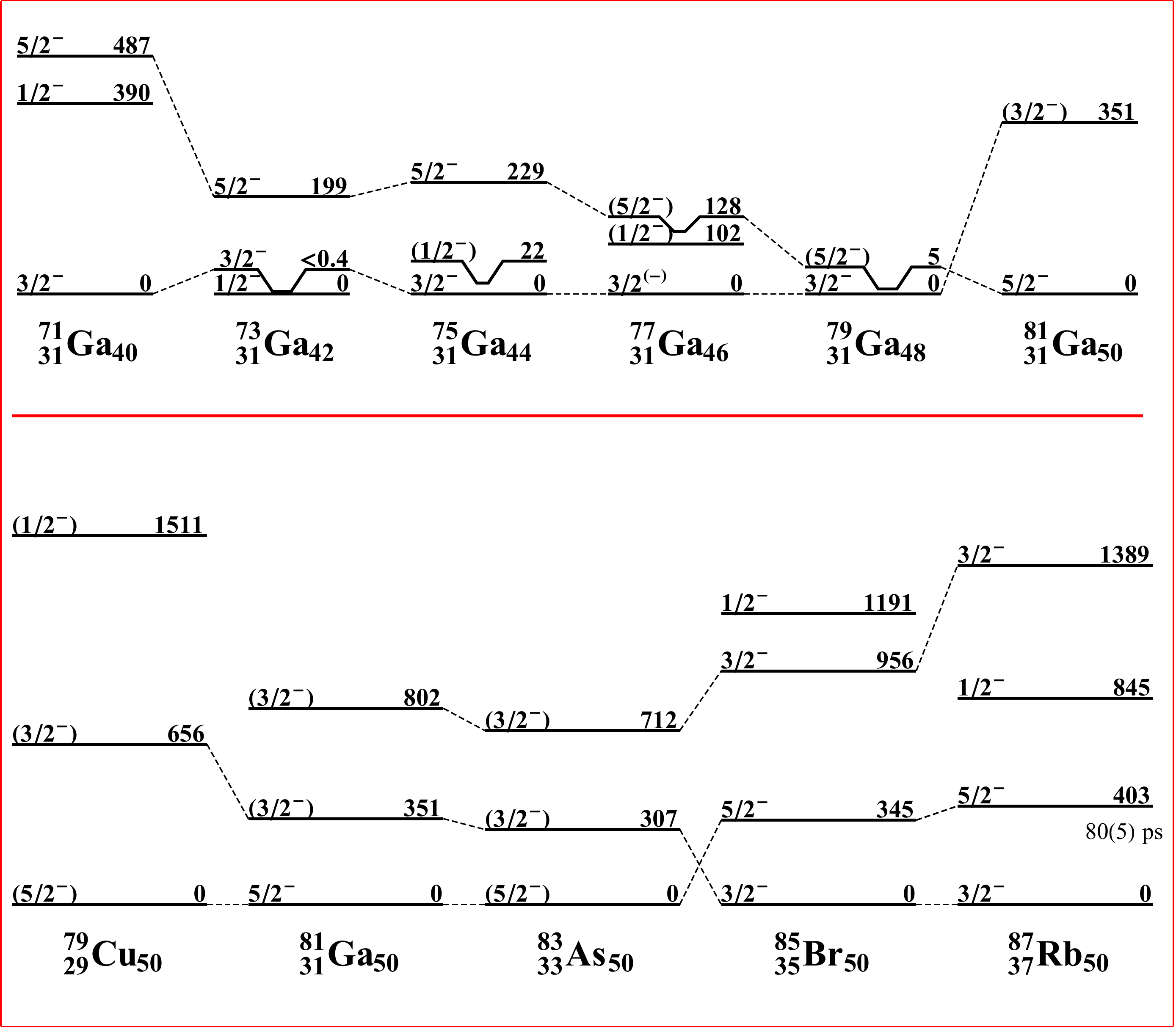}}
	\caption{\label{N50_isotones_Ga} Top: level systematics of $N=40-50$ Ga isotopes. Bottom: $N=50$ isotones. The low-lying levels of Ga isotopes are taken from references \cite{zoller70,diriken10,ekstrom86,phoff81}, except for $^{81}$Ga from this work, while the structure of the $N=50$ isotones is based on \cite{olivier17,winger88,zendel80,wohn73}.}
\end{figure*}

According to the measured $B(XL)$ values, both the 351- and 452-keV transitions are consistent with having a predominant $M1$ character as in the case of the 345-keV transition in the $N=50$ $^{85}$Br isotone \cite{NDS85} connecting the well-established $3/2^{-}$ and $5/2^{-}$ states, as shown in Fig.~\ref{N50_isotones_Ga}. Based on systematics, an $M1$ multipolarity is also suggested in \cite{sahin12} for the 307-keV transition connecting the tentatively assigned $(3/2^{-})$ first excited state and $(5/2^{-})$ ground state in $^{83}$As, which was also measured in \cite{porquet11}.

\section{Shell-model calculations}

Large-scale shell-model calculations of nuclear states of $^{81}$Ga have been performed. Two state-of-the-art effective  interactions were implemented into the NuShellX@MSU \cite{lisetskiy04} and ANTOINE \cite{Caurier2005} codes. The first interaction, labelled JUN45, was developed by Honma \textit{et al.} in 2009 \cite{honma09} and it was focused in the \textit{pf} shell with a $^{56}$Ni core and contains the $1p_{3/2}$, $0f_{5/2}$, $1p_{1/2}$ and $0g_{9/2}$ single-particle orbits. 
The interaction reproduces the experimental data of low-lying states in the $N=49$ isotones, Ge isotopes near $N=40$, and $N=Z$ nuclei with $A=64\sim70$, but the valence space may not contain all the degrees of freedom necessary to account for all the features of the nuclear structure of the region \cite{honma09}.

Another effective interaction, called jj44b, which made successful predictions for nuclei near $^{78}$Ni, was created in 2004 by Lisetskiy \textit{et al.} \cite{lisetskiy04}. It was constructed with a $^{56}$Ni core for the neutron space and a $^{78}$Ni core for the proton space. The Hamiltonian was also based on the Bonn-C $NN$ potential including four single-particle energies and 65 $T=1$ two-body matrix elements. The interaction was later updated \cite{verney07} 
to better describe the structure of $^{81}$Ga and it has been shown to reproduce the properties of the heavier isotopes of Ga \cite{srivastava12}.
Here we employ the original jj44b interaction.

\begin{figure*}[!]
	\centerline{\includegraphics[scale=0.45]{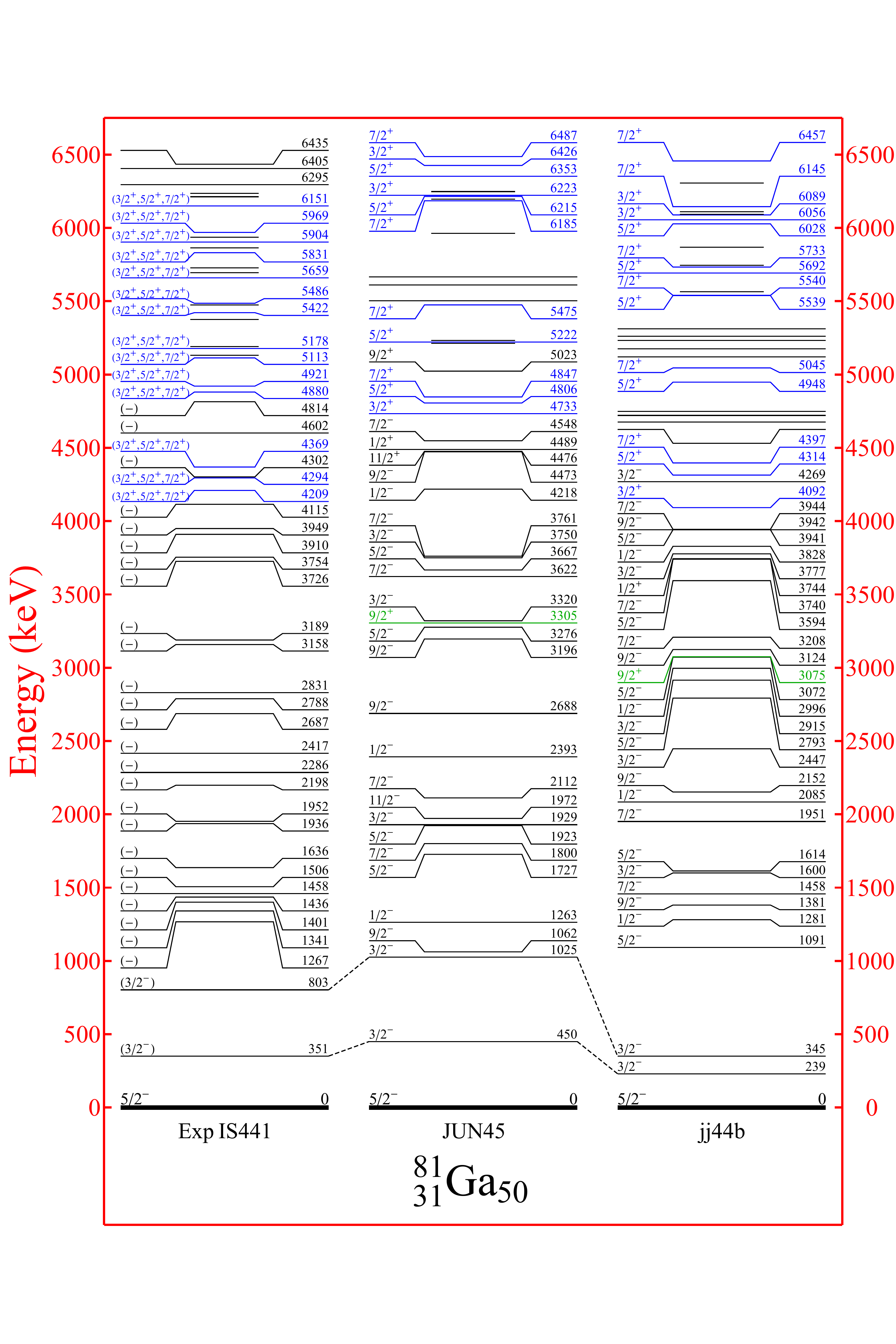}}
	\caption{\label{levels_SM} Shell-model calculations with the JUN45 and jj44b interactions for $^{81}$Ga compared to the experimental data measured in this work. The experimental spins were tentatively placed based on the $\beta$-feeding considerations. The positive-parity states are marked in blue, except for the calculated 9/2$^+$ state, which is highlighted in green.} 
\end{figure*}

The energy levels of $^{81}$Ga obtained with the JUN45 and jj44b interactions are compared to our experimental results in Fig.~\ref{levels_SM}.
The calculations using the JUN45 interaction achieve a good agreement with the experiment for the excitation energy of the low-lying states, but tend to overestimate the energy of the negative-parity levels in the 1$-$2~MeV region. On the other hand the jj44b interaction fails to reproduce the energy
of the second-excited state, but achieves a better description of the level density in the 1$-$2~MeV region. 

Both sets of calculations give the lowest-lying positive-parity state as a spin and parity of 9/2$^+$ that will arise from the $\pi(g_{9/2})$ configuration, at around 3 MeV. The higher-lying positive-parity states obtained from the calculations in this restricted model space must arise from the coupling of a $\pi(g_{9/2})$ proton to a proton pair in the negative-parity orbitals. These states cannot be related to the experimentally observed ones, since the latter
should have a neutron intruder nature (1p-1h neutron configurations) in order to be connected to the $^{81}$Zn ground state via Gamow-Teller transitions. 

The occupation probabilities predicted with both interactions for the lowest-lying states are summarized in Table~\ref{tab:occupations}. Both sets of shell-model calculations are able to properly reproduce the ground-state spin-parity to be 5/2$^-$, in agreement with the experimental value \cite{cheal10}. The proton occupation probability for the $\pi(f_{5/2})^3$ configuration is 79\% with JUN45 and 75\% with jj44b. A spin-parity of 3/2$^-$ is calculated for the first and the second excited states with both interactions, although their energies differ considerably. According to the calculations using the jj44b interaction, the 351-keV level has a single-particle $\pi(p_{3/2})$ character, with a large occupation value for the $\pi(f_{5/2})^2$$\otimes$$\pi(p_{3/2})$ configuration, whereas the 802-keV level has a preferred $\pi(f_{5/2})^3$ configuration. The calculations for $^{81}$Ga using the PFSDG-U interaction \cite{nowacki16} reported in \cite{dudouet19} support this description. 

\begin{table}
\centering
\caption{Occupation probabilities for the proton configurations obtained with the JUN45 and jj44b interactions for the first
three states in $^{81}$Ga.}
\label{tab:occupations}
\begin{tabular}{cccccc}
\hline
$\bm{E_{level}}$ & $\bm{J^{\pi}}$ & \multicolumn{2}{c}{\textbf{JUN45}} & \multicolumn{2}{c}{\textbf{jj44b}}\\ 
 \textbf{(keV)}  &                &  $\pi(f_{5/2})^3$ & $\pi(f_{5/2})^2p_{3/2}$ &  $\pi(f_{5/2})^3$ & $\pi(f_{5/2})^2p_{3/2}$ \\
\hline
0       & 5/2$^{-}$   &  79\% & 4\%  & 75\% & 2\%\\
351     & (3/2$^{-}$) &  38\% & 51\% & 5\%  & 78\%\\
802     & (3/2$^{-}$) &  48\% & 41\% & 76\% & 5\%\\ 
\hline
\end{tabular}
\end{table}

The calculated reduced transition probabilities for the $\gamma$ rays de-exciting the lowest-lying excited states are given in Table~\ref{transition_prob} and compared with the measured values based on the experimental lifetimes and branching ratios. Only parity-conserving transitions of the lowest multipolarities are considered due to the negative parity nature of the low-lying states. Here effective charges of $e_{\pi}=+1.5e$ and $e_{\nu}=+1.1e$ were used, along with a quenched $g$ factor of $0.7g_{s,free}$. 
The harmonic oscillator potential used was $41A^{-1/3}$ MeV, as recommended in \cite{vin10}, which was found to better reproduce the transition rates in this region.

\section{Discussion}

\subsection{Ground-state feeding of \texorpdfstring{$^{\bm{81}}$}{81Ga}Ga and ground-state spin-parity of \texorpdfstring{$^{\bm{81}}$}{81Zn}Zn}

The structure of the nuclei immediately north of $^{78}$Ni is defined by the ordering and occupation probabilities of the $g_{9/2}$, $d_{5/2}$, $s_{1/2}$, $g_{7/2}$, and $d_{3/2}$ neutron orbitals, and the $f_{7/2}$, $p_{3/2}$, $p_{1/2}$, and $f_{5/2}$ proton orbitals. For the Cu ($Z=29$) isotopic chain, an inversion of the ordering of the $p_{3/2}$ and $f_{5/2}$ states above $^{75}$Cu has been observed as the neutron $g_{9/2}$ orbital is being filled. This has been interpreted as the effect of the monopole neutron-proton tensor interaction \cite{otsuka05}.

The $^{81}$Ga ground-state spin-parity has been established as 5/2$^-$ from collinear laser spectroscopy performed at ISOLDE \cite{cheal10}.
For the $^{81}$Zn g.s., positive-parity states with a single-particle character, $\nu s_{1/2}$ or $\nu d_{5/2}$, have been proposed, leading 
to either 1/2$^+$ spin-parity assignment, matching the extrapolation of the 1/2 state energies in the region \cite{verney07}, or 5/2$^+$, which is consistent with the systematics of $N=51$ isotones. 
Beta-decay transitions from $^{81}$Zn to low-lying states in $^{81}$Ga are expected to proceed via forbidden transitions, since the Gamow-Teller operator will populate daughter states at much higher energy. 
Using the systematics for forbidden decays \cite{logft_review} and based on the firm spin-parity assignment 5/2$^-$ for the ground state of $^{81}$Ga, two options are possible: a first-forbidden decay from the 5/2$^+$ to 5/2$^-$ states with log$ft>5.9$ or a first-forbidden unique decay from 1/2$^+$ to 5/2$^-$, more hindered and with log$ft>7.5$. 

In \cite{pad10} Padgett \textit{et al.} ruled out the previous suggestion of 1/2$^+$ \cite{verney07} for the ground state spin-parity of $^{81}$Zn. Instead a 5/2$^+$ assignment for the $^{81}$Zn ground state was proposed \cite{pad10} based on the apparent $\beta$-decay feeding to the $^{81}$Ga 5/2$^-$ ground state. 
Our experimental data yield a $\beta$ ground-state feeding compatible with zero, with a log$ft$ larger than 6.8. We note that high-energy transitions can still be missed in our detection set-up 
and that, on the contrary, some of the high-energy transitions that are unambiguously identified as belonging in $^{81}$Ga have been tentatively placed in the level scheme as directly feeding the ground state based on the lack of observed coincidences. Thus the experimental value needs to be taken with caution. 
Nonetheless, the observed negligible direct $\beta$ feeding to the ground state is in contrast to the previous experiment \cite{pad10}. This is because many weak $\gamma$ transitions de-populating high-lying states in $^{81}$Ga have been added in our study, which has a dramatic effect on the ground-state $\gamma$ feeding intensity and decreases the apparent direct ground-state $\beta$ branching from the previous 52\% to our $\leq$2.4\%.
It is therefore very risky to base spin-parity assignments on the apparent $\beta$-feeding. From our measurement none of the possible spin-parity assignments of $^{81}$Zn can be ruled out, since a first forbidden unique 1/2$^+$ to 5/2$^-$ transition, and thus a 1/2$^+$ $^{81}$Zn ground-state spin-parity, is still possible. 

In any case, the role of first-forbidden transitions to low-lying negative-parity states in $^{81}$Ga is not as relevant as previosuly proposed. A large fraction of the beta-decay population may still proceed to higher-lying positive-parity states via allowed transitions. This is ratified by the sizeable $\beta$-delayed neutron emission probability, which points to $\beta$ feeding via GT transitions to high-energy levels above the neutron separation energy in $^{81}$Ga.

\subsection{Low-lying structure of \texorpdfstring{$^{\bm{81}}$}{81Ga}Ga}


In our experiment we have measured the first-excited 351-keV state half-life to be $T_{1/2}~=~60(10)$ ps. Assuming that the transition connects states of negative parity, the $B(M1)$ rate for pure $M1$ multipolarity is ${8.5(14)}\times{10^{-3}}$ W.u., whereas the pure $E2$ rate would be a very collective 85(14) W.u. The reduced transition probabilities thus point to a retarded $M1$ transition which is consistent with the results of the shell-model calculations (Table \ref{transition_prob} and the systematics in \cite{endt1979}). This suggests a ${\pi}p_{3/2}$ dominant configuration for the 351-keV level (see Table \ref{tab:occupations}). Owing to this fact, the 351-keV $\gamma$ ray would be slightly hindered due to the $l$-forbidden character of a ${\pi}p_{3/2}$ to ${\pi}f_{5/2}$ transition. 

A very similar structure is found in odd $N=50$ nuclei with $Z>28$, as shown in Fig.~\ref{N50_isotones_Ga}, in particular for the neighboring isotone $^{83}$As. For the $Z=37$ $^{87}$Rb isotone, the 3/2$^{-}$ and 5/2$^{-}$ levels are already reversed, the ground state having a spin of 3/2$^{-}$. The 403-keV 5/2$^{-}$ level has a very similar half-life of $T_{1/2}=80(5)$ ps \cite{nds87} to the 351-keV one in $^{81}$Ga, which gives a $B(M1)$ to within a factor of 2 for the 403-keV transition to the ground state compared to the 351-keV transition in $^{81}$Ga. The dominant ${\pi}p_{3/2}$ single-particle configuration of the 351-keV level is in accord with a narrow proton gap of the order of 500 keV between the $f_{5/2}$ and $p_{3/2}$ orbitals, as predicted for $^{79}$Cu by shell-model calculations \cite{smir04,vin10}, but at odds with what was claimed in \cite{daugas10}. 

For the second excited state at 802 keV, the calculations, especially those with the JUN45 interaction (which achieve a better agreement with the experimental excitation energies) show a strong admixture of the $\pi(f_{5/2})^3$ cluster configuration and the $\pi(f_{5/2})^2 p_{3/2}$ one. This gives rise to a 3/2$^-$ spin-parity. Our measured half-life for this level is consistent with a 3/2$^-$ assignment, and an $M1$ 452-keV de-exciting transition, whereas for the 802-keV transition the experimental result allows for a possible $E2$ component. 

Both of the 351- and 802-keV levels are fed from positive-parity high-energy levels. These may be characterized by the occupation of ${\pi}g_{7/2}$, ${\pi}d_{5/2}$ and ${\pi}d_{3/2}$ proton single-particle states, or by the coupling of proton orbitals to neutron particle-hole states (thus requiring breaking of neutron pairs 
across the $N=50$ gap). The de-excitations from these high-lying states, which likely have spins between $1/2^{+}$ and $7/2^{+}$, to $3/2^{-}$ 351- and 802-keV states take place via $E1$ high-energy transitions with energies higher than 3.8 MeV. 
In view of the 3.5-MeV $N=50$ energy gap measured for $^{81}$Ga by Hakala \textit{et al.} \cite{hakala08} the 3859-keV transition which connects the $\beta$-fed 4209-keV level to the 351-keV state gives a rough estimate of the $N=50$ energy gap from our data.      

Nine other excited states are experimentally found in $^{81}$Ga below 2 MeV. The calculations reproduce the level density of these negative parity states. In a simplistic model where a quasiparticle is coupled to the $^{80}$Zn core, four states arising from the coupling of the $\pi(p_{3/2})$ orbit to the 2$^+$ core in $^{80}$Zn, would have spins of 1/2$^-$, 3/2$^-$, 5/2$^-$, and 7/2$^-$, whereas the $\pi(f_{5/2})$ coupling to the 2$^+$ level will give rise to the five states with spins ranging from 1/2$^-$ to 9/2$^-$. An alternative description based on a $\pi(f_{5/2})^3$ cluster configuration provides a similar picture. The $\pi(f_{5/2})^3$ configuration yields 3/2$^-$, 5/2$^-$ and 9/2$^-$ spins, with the 9/2$^-$ found at higher energies, and their couplings to the 2$^+$ of $^{78}$Ni in this case will provide the observed levels. 
It is worth noting that both theoretical calculations reproduce rather well the excitation energy of the 9/2$^-$ level at 1341.0 keV belonging to the $\pi(f_{5/2})^3$ configuration.
This is consistent with the calculations presented in Ref. \cite{dudouet19}.

A high density of levels in the region from 1 to 2 MeV can be observed as well in the level scheme of $^{83}$As \cite{winger88},
with striking similarity to that of $^{81}$Ga. The level scheme of $^{85}$Br \cite{NDS85}, populated by the $\beta$ decay of $^{85}$Se has a very similar structure too. Out of these levels in $^{81}$Ga, the 1936-keV state is strongly populated from the higher lying 4295-keV positive-parity state. We have measured a 21-ps upper half-life limit for the former, which does not allow us to unambiguously identify the multipolarity of the de-populating 1936- and 1585-keV transitions. However, the decay pattern to the ground and first-excited levels, and the direct feeding from positive-parity states, makes a spin-parity assignment of 3/2$^-$ or 5/2$^-$ likely for this state. 

\subsection{Positive-parity states}

As mentioned above, the lowest positive-parity state is the $9/2^{+}$ one predicted at energies close to 3.0 and 3.3 MeV, depending on the interaction. This state has an expected main ${\pi}g_{9/2}$ configuration and would not be directly populated by the $^{81}$Zn $\beta$ decay from a $1/2^{+}$ ground state, and would have a limited feeding from a $5/2^{+}$ g.s., yielding a high log$ft$ value. 
Although the single-particle ${\nu}g_{7/2}$ orbit is shown to be at higher energy \cite{sahin12}, any admixture of a ${\nu}g_{7/2}$ component in the $^{81}$Zn g.s. wave function would lead to an enhancement of allowed Gamow-Teller (GT) $\beta$ transtions to the ${\pi}g_{9/2}$ orbit. 
In any case, indirect population of the $9/2^{+}$ state in $^{81}$Ga should be possible. The systematics near $A=81$ suggests a long half-life for this state due to the $M2$ behavior of the $\gamma$ transition which would connect it to the $5/2^{-}$ ground state. No such long lifetime, nor decay to lower energy $7/2^{-}$, $9/2^{-}$ levels, could be observed in our measurement.

The allowed GT $\beta$ decay from the $^{81}$Zn ground-state neutron $\nu(d_{5/2})$ or $\nu(s_{1/2})$ configuration populates high-energy states in the $^{81}$Ga daughter, since there are no low-lying positive-parity states available. The positive states would have to originate from the coupling of the odd proton orbitals ($p_{3/2}$, $f_{5/2}$ and $p_{1/2}$) to neutron particle-hole states, therefore implying the breaking of a neutron pair inside the $N=50$ shell. These cross-shell states arising from the excitation of the $^{78}$Ni core give an idea of the magnitude of the $N=50$ shell gap, as discussed by Winger \textit{et al.} \cite{winger88} in the $\beta$ decay of $^{83}$Ge to $^{83}$As, and Padgett and co-workers for $^{81}$Ga \cite{pad10}. 

The GT $\beta$ decays to these core-excited states must arise from the decay of neutrons in $^{81}$Zn in the $f$ and $p$ orbitals, which are strongly bound. Due to the reduced energy window the $\beta$ feeding would be reduced, but, in spite of the Fermi factor, these GT decays may still be favoured compared to the first-forbidden decays to low-lying negative-parity states. The large P$_n$ value measured for $^{81}$Zn suggests a significant role of such allowed $\beta$ transitions to high-lying states above the neutron separation energy in $^{81}$Ga. 
Several levels with low apparent log$ft$ values can be identified in the 4 -- 5 MeV energy range. In our work we observe strong direct population to the levels at 4209, 4295, 4369, 4880, 4921, 5178 and 5422 keV, and to some others at higher energies. These states are not included in our shell-model calculations due to the restricted model spaces. 
Assuming a $1/2^{+}$ or $5/2^{+}$ g.s. for $^{81}$Zn, positive-parity assignments for these levels with 1/2, 3/2, 5/2, and 7/2 spin values can be made. 
An identical situation can be observed in the $N=50$ isotones $^{83}$As \cite{winger88} and $^{85}$Br \cite{NDS85}, populated following the $\beta$ decay of $^{83}$Ge and $^{85}$Se, respectively.

\section{Summary and conclusions}

The high purity and intensity of the Zn beams delivered by the ISOLDE facility at CERN have made it possible to obtain about ten-fold higher statistics than previous studies \cite{pad10}. The level scheme of the semi-magic $N=50$ nucleus $^{81}$Ga has been significantly expanded with 47 new levels and 70 $\gamma$ transitions in the energy range up to 6.5~MeV. Most of these levels are very close to the neutron separation energy. The 290(4)-ms half-life of $^{81}$Zn measured in this work is in good agreement with the literature \cite{pad10,xu14}.

The direct $\beta$ feeding to the $^{81}$Ga ground state measured in our experiment is negligible within the error bars, and much lower than proposed previously; it is thus compatible with both $5/2^{+}$ and $1/2^{+}$ assignments for the $^{81}$Zn ground state. We could not identify the $9/2^{+}$ state seen in other $N=50$ isotones and also predicted by our shell-model calculations to lie at around 3 MeV. We have measured a $\beta$-delayed neutron emission probability value of 23(4)\% for the decay of $^{81}$Zn. This is more precise but also consistent with 30(13)\% measured by Hosmer \textit{et al.} {\cite{hosmer10}}, but two-sigma away from the recent value reported by Padgett and co-workers of 12(4)\% \cite{pad10}.

The level scheme of $^{80}$Ga populated following the $\beta$-delayed neutron emission from $^{81}$Zn was constructed for the first time and it is in agreement with that described  in {\cite{RAZ14}} from the $\beta$ decay of $^{80}$Ga, including the low-lying 22-keV isomer. Our measurements also confirm the existence of the 708-keV isomer with an 18.3(13)-ns half-life.

We have measured the half-life of the first excited state in $^{81}$Ga to be $T_{1/2}=60(10)$ ps, which indicates an \textit{l}-forbidden $M1$ transition of 351 keV to the 5/2$^{-}$ ground state. This in turn points to a transition between states with main $\pi{p_{3/2}}$ and $\pi{f_{5/2}}$ configurations. This is supported by both the $N=50$ systematics and by our shell-model calculations, where the dominant occupations for the ground and first-excited states are found, and in agreement with earlier findings \cite{verney07}. The calculated occupation probability and our experimental results suggest a main $\pi(f_{5/2})^2$$\otimes$$\pi(p_{3/2})^1$ configuration for the first excited state of $^{81}$Ga. The calculated transition rate supports this assignment too. For the second excited state a half-life of 23(16) ps is measured. This value provides $B(M1)=8(6)\times{10^{-3}}$ W.u. and $B(E2)=51(35)$ W.u. (Table~\ref{transition_prob}) reduced probabilities which, together with the shell-model results, allows us to propose a $\pi({f_{5/2}})^3$ cluster configuration and a $3/2^{-}$ spin-parity assignment for this state. 

A high density of negative-parity levels can be observed in the region from 1 to 2 MeV of the level scheme of $^{81}$Ga. This is consistent with $\pi(p_{3/2})$ and $\pi(f_{5/2})$ single-particle states coupled to the 2$^+$ core in $^{80}$Zn, and it is well reproduced by the shell-model calculations. These states will be of negative parity and should be populated by first-forbidden transitions if they are directly $\beta$ fed. The level scheme of the $N=50$ isotone $^{83}$As {\cite{winger88}} also shows a density of levels around 1400 keV much like that of $^{81}$Ga. A similar structure is found in the $N=50$ $^{85}$Br isotone populated by the $\beta$ decay of $^{85}$Se {\cite{NDS85}}. The situation changes beyond 5 MeV where we observe several states with sizeable apparent $\beta$ feeding, which should arise from allowed transitions from the $^{81}$Zn positive-parity ground state. They can be interpreted as neutron particle-hole excitations from the $^{78}$Ni core. 


\begin{acknowledgments}

The authors would like to express profound recognition to the late Prof. Henryk Mach, who pioneered the $\beta\gamma\gamma$(t) fast timing method and its application to exotic nuclei around $^{78}$Ni. Henryk was a true friend and an inspirational character. His premature passing away is a major loss for our community both from the human and scientific point of view.

This work was supported by the Spanish MINECO through the FPA2015-65035-P and RTI2018-098868-B-I00 projects, by the US DoE Grant No. 
DE-FG02-94ER40834, by the German BMBF Grant 05P19PKFNA, and Grupo de F\'{\i}sica Nuclear (GFN) at UCM. The support by the European Union Seventh Framework through ENSAR (contract no. 262010) and ISOLDE (CERN) Collaboration is acknowledged. V.P. acknowledges support by the Spanish FPI-BES-2011-045931 grant. Fast-timing electronics were provided by the Fast Timing Collaboration and MASTICON. Figures \ref{LevelScheme_81Ga_low}, \ref{LevelScheme_81Ga_high} and \ref{level_scheme_80Ga} were created using the LevelScheme scientific figure preparation system \cite{caprio2005}.
 
\end{acknowledgments}


\providecommand{\noopsort}[1]{}\providecommand{\singleletter}[1]{#1}%

\end{document}